\documentclass[
	reprint,
	groupedaddress,
	amsmath,amssymb,
	prb,
	floatfix,
	superscriptaddress,
	twocolumn,
	aps,
]{revtex4-2}
\usepackage[utf8]{inputenc} %
\usepackage[T1]{fontenc}    %
\usepackage{helvet}         %
\usepackage{mathpazo}       %
\usepackage{microtype}      %
\usepackage{amsmath}
\usepackage{bm}
\usepackage{amsfonts}
\usepackage{amssymb}
\usepackage{amsthm}
\usepackage[caption=false]{subfig}
\usepackage{graphicx}
\usepackage{hyperref}
\usepackage{cleveref}
\usepackage[T1]{fontenc}
\usepackage{multirow}
\usepackage{lipsum} %
\usepackage{color,soul}
\usepackage[export]{adjustbox}
\usepackage{natbib}
\usepackage{booktabs}
\usepackage{todonotes}

\usepackage{xr}
\begin{document}

\title{Synthesizability, hardness, and stacking order in multicomponent transition metal carbides from machine-learned potentials}

\author{Xin Liu}
\email{xin.liu@epfl.ch}
\affiliation{Laboratory of materials design and simulation (MADES), Institute of Materials, \'{E}cole Polytechnique F\'{e}d\'{e}rale de Lausanne}

\author{Anirudh Raju Natarajan}
\email{anirudh.natarajan@epfl.ch}
\affiliation{Laboratory of materials design and simulation (MADES), Institute of Materials, \'{E}cole Polytechnique F\'{e}d\'{e}rale de Lausanne}
\affiliation{National Centre for Computational Design and Discovery of Novel Materials (MARVEL), \'{E}cole Polytechnique F\'{e}d\'{e}rale de Lausanne}

\begin{abstract}
  Multicomponent transition metal carbides are promising for extreme-environment applications, but identifying compositions that are both synthesizable and hard remains challenging. We fine-tune the MACE machine-learned interatomic potential on approximately 28,000 density functional theory calculations spanning the composition space of groups 4--6 transition metals and carbon to predict the thermodynamic stability and elastic properties of multicomponent carbides. The fine-tuned model achieves formation energy errors of $\sim$ 10\,meV/atom for thermodynamically relevant structures with only 20\% of the training data. We screen over 1500 equiatomic compositions across rocksalt, hexagonal, and hcp prototypes, combining free energy models with elasticity-based hardness surrogates. Synthesizability predictions at 1500\textdegree C agree well with experimental reports for both single-phase and multiphase carbides. The group number of the constituent metals governs both stability and hardness. Free energy contributions from short-range order are small, typically a few meV/atom, indicating that a perfectly disordered solid solution provides a reasonable approximation for high-throughput screening. For compositions mixing group 4/5 and group 6 metals, we identify a new family of stacking-ordered phases with formation energies well below those of disordered rocksalt or hexagonal structures. DFT calculations corroborate these predictions and suggest that stacking-ordered phases should be experimentally accessible in multicomponent carbides. This study provides a framework for screening synthesizable multicomponent materials with target properties, identifies promising carbide compositions across the full nine-component space, and reveals a new class of stacking-ordered carbides accessible only in multicomponent compositions.   
\end{abstract}

\maketitle

\section{Introduction}
\label{sec:introduction}

Transition metal carbides (TMCs) are candidates for extreme environments due to their high melting points, favorable mechanical properties, and high thermal conductivity. Recent studies have sought synthesizable carbides for applications including hypersonic flight, nuclear reactors, and corrosive environments, with multicomponent TMCs (sometimes referred to as high-entropy carbides) attracting increased attention  \cite{Wyatt:2024aa, WILLIAMS197157, Wang2020, zhang_review_2019, prats_transition_2024, BIE2026153140}. Researchers have expanded into multicomponent composition space given the vast number of potentially synthesizable materials and the possibility of achieving properties that surpass those of binary or ternary counterparts. However, finding promising materials within this large design space requires reliable predictions of synthesizability and material properties.

\begin{figure}[!ht]
  \centering
  \includegraphics[width=0.45\textwidth]{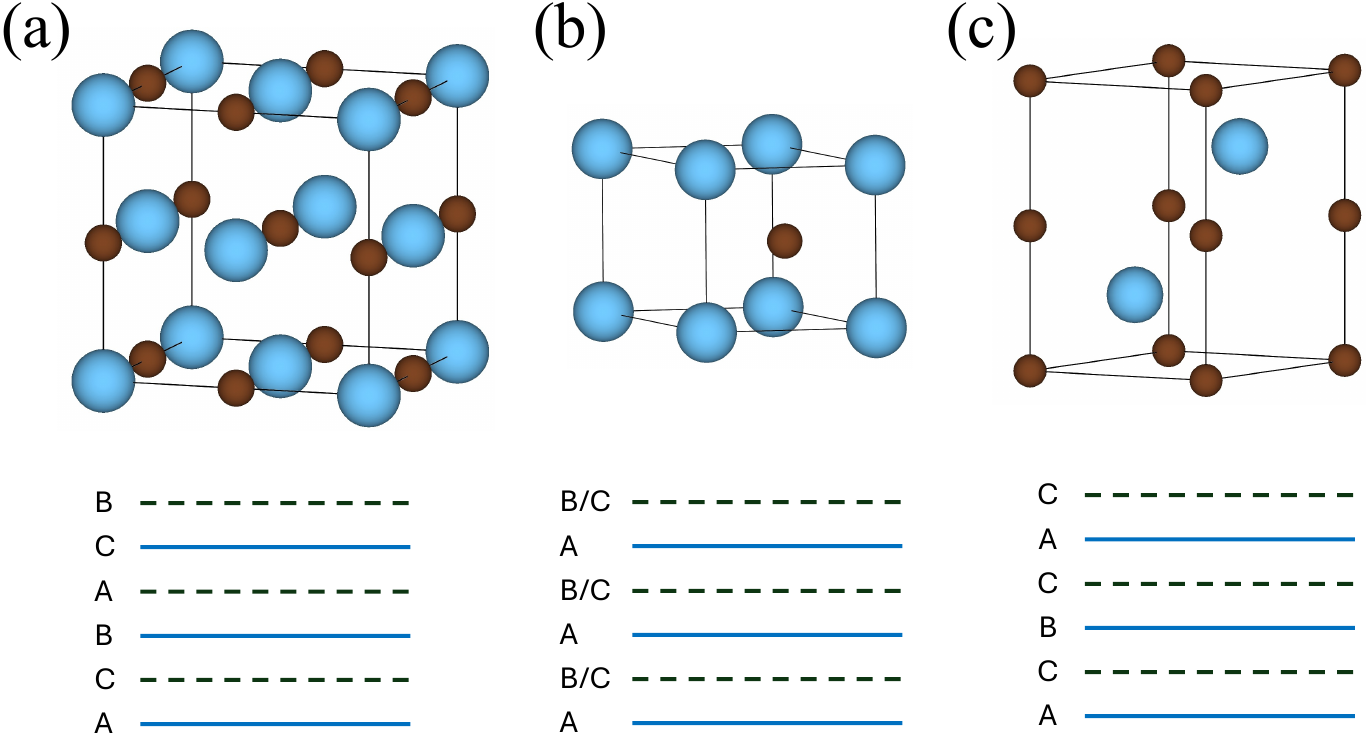}
  \caption{Parent crystal structures of groups 4--6 transition metal carbides. (a) Rocksalt (Fm$\bar{3}$m). (b) Hexagonal WC-type (P$\bar{6}$m2). (c) HCP (P$6_3$/mmc). Light-blue spheres represent metal atoms and brown spheres represent carbon or vacancy sites. The lower half of each panel shows the stacking of metal and carbon layers along the $c$ axis (hcp and hexagonal) or a [111] direction (rocksalt). Solid lines denote metal layers, dashed lines denote carbon layers, and letters indicate the relative stacking (A, B, C) of each triangular layer.}
  \label{fig:structure_prototypes}
\end{figure}

The possibility of TMCs adopting multiple structural prototypes further complicates the design of these materials. Carbides with a 1:1 metal-to-carbon ratio formed from group 4 and group 5 elements crystallize in the rocksalt structure (\cref{fig:structure_prototypes}a), where metal atoms form a face-centered cubic lattice with ABC-type stacking and carbon atoms occupy octahedral interstices. Group 6 carbides such as molybdenum and tungsten carbide adopt a different arrangement, with metal atoms in a simple hexagonal lattice (\cref{fig:structure_prototypes}b) stacked in an AA-type sequence and carbon occupying octahedral sites between the metal planes \cite{Gunda2018}. Beyond these stoichiometric structures, most metals in groups 4--6 tolerate substantial carbon off-stoichiometry, forming carbon-vacancy orderings at low temperature or carbon-vacancy disorder at elevated temperatures. Reducing the carbon content below $x_{C}=0.5$ can stabilize hexagonal close-packed (hcp) stacking of the metal sublattice (\cref{fig:structure_prototypes}c), particularly in the presence of group 4 elements such as titanium, zirconium, or hafnium.

Several recent studies have screened multicomponent composition spaces to identify synthesizable TMCs with a rocksalt crystal structure. These efforts used first-principles calculations to assess the stability of disordered rocksalt phases through descriptors including entropy-forming ability \cite{sarker_high-entropy_2018}, mixed enthalpy-entropy metrics \cite{dey_mixed_2024}, disordered enthalpy-entropy descriptors \cite{divilov_disordered_2024}, lattice-distortion measures \cite{Hedman2022}, and formation energies combined with approximate free energy models \cite{qureshi_predictive_2025}. While experiments have validated predictions from these approaches, much of the design space remains unexplored. Most studies focus on synthesizability without assessing properties relevant to extreme environments, such as strength or hardness. Additionally, existing work typically restricts investigation to quinary TMCs and considers disorder only over the rocksalt structure, neglecting other viable structures such as hcp and hexagonal carbides.

In this study, we fine-tune a universal machine-learned interatomic potential on high-throughput electronic structure calculations to predict the thermodynamics and mechanical properties of multicomponent TMCs across the composition space of groups 4--6 elements. We combine this potential with free energy models and elasticity-based hardness surrogates to screen the nine-component composition space for disordered rocksalt, hexagonal, and hcp carbides that are both synthesizable and hard. The rocksalt structure accommodates the greatest degree of chemical disorder, while hexagonal and hcp structures tolerate disorder only over limited composition ranges. Across all structures, the group number of the constituent metals governs both stability and hardness, and short-range order has limited impact on either property. For compositions mixing group 4/5 and group 6 metals, the fine-tuned potential predicts a new family of stacking-ordered phases, analogous to those in lithium, with metal sublattices adopting various stacking sequences and carbon distributed over octahedral interstices. These stacking variants are accessible only in multicomponent TMCs, a prediction we validate through additional DFT calculations. This study provides insight into the interplay between synthesizability and hardness, reveals a new class of stacking-ordered carbide prototypes, and presents a tailored interatomic potential for predicting properties of groups 4--6 carbides.

\section{Methods}
\label{sec:methods}

Assessing the finite-temperature thermodynamic stability of multicomponent disordered carbides, along with properties such as elastic moduli and hardness, requires an accurate atomistic model and finite-temperature simulations grounded in statistical mechanics. We trained an atomistic model by fine-tuning a universal machine-learned interatomic potential (MLIP) on a large dataset of electronic structure calculations. We then combined this potential with both approximate free energy models and more rigorous statistical mechanics methods to assess the effects of temperature and short-range order on phase stability and to predict the hardness of disordered multicomponent TMCs.

We fine-tuned an MLIP based on the MACE architecture \cite{Batatia2022Design,Batatia2022mace} to predict the atomistic properties of multicomponent TMCs. The MACE architecture extends the atomic cluster expansion framework \cite{drautz_atomic_2019}, and previous benchmarks demonstrate its applicability across a broad range of materials \cite{Kovacs2023}. Fine-tuning of the \texttt{MACE-MPA-0} foundation model used a dataset of approximately 28,000 density functional theory (DFT) calculations. These calculations were performed with the Vienna Ab initio Simulation Package (VASP) \cite{Kresse199611169}, employing the Perdew–Burke–Ernzerhof exchange-correlation functional \cite{Perdew1996} and projector augmented-wave pseudopotentials \cite{Kresse19991758}. A plane-wave cutoff of 550\,eV and a $\Gamma$-centered k-point mesh with 0.1\,\r{A}$^{-1}$ spacing were applied. Partial occupancies were treated with the second-order Methfessel–Paxton scheme \cite{Methfessel19893616} and a smearing width of 0.1\,eV. Spin polarization was included for systems containing Cr and V, with magnetic moments initialized at 3\,$\mu_B$.

The dataset includes structures composed of nine transition metals (Ti, Zr, Hf, V, Nb, Ta, Cr, Mo, W) and carbon. To ensure broad chemical and structural coverage, the dataset spans chemical decorations of the three structural prototypes (\cref{fig:structure_prototypes}) with several carbon concentrations, as well as chemical orderings of the metallic elements over the bcc crystal structure. The training set also includes structural perturbations of low-energy orderings to capture elastic constants, generalized stacking fault energies, decohesion, and surface energies. This range of structures enables the model to accurately describe the effects of chemical ordering on thermodynamic stability, elasticity, and surface energetics.

We fine-tuned the MLIP on both DFT-computed energies and forces, withholding 5\% of the data for validation. Training followed a two-stage protocol. In the first stage, energy and force contributions were weighted equally over 65 epochs. In the second stage, the energy weighting was increased to 5 to reduce energy prediction errors. We terminated training early if no validation error improvement was observed over 20 consecutive epochs during the second stage.

We assessed finite-temperature phase stability by comparing free energies of all relevant phases. To determine the thermodynamic stability of a carbide at finite temperature, we first computed its per-atom formation energy relative to elemental reference states:
\begin{equation}
  \Delta e_f = \frac{1}{\sum_i n_i}\left[E_{tot} - \sum_i n_i e_i \right]
  \label{eq:free_energy_per_atom}
\end{equation}
where $E_{tot}$ is the total energy of the carbide, $n_i$ is the number of atoms of element $i$, and $e_i$ is the reference energy of element $i$ in its stable ground-state structure. Reference states correspond to the lowest-energy structures in the training database for the pure elements.

We evaluated the free energy of the disordered phase, $\Delta g_{disorder}$, using two approximations. Both neglect vibrational entropy, whose contributions are often similar across competing phases in multicomponent materials. When entropy arising from chemical disorder dominates phase stability, the free energy can be computed as \cite{Helmholtz1882, muller_first-principles_2024}:
\begin{equation}
    \Delta g_{disorder} \left(T\right)  = \frac{T}{T_0}\Delta g_{disorder} \left(T_0\right) + T\int_{T_0}^T -\frac{\langle \Delta e_f \rangle}{\tau^2}d\tau
\label{eq:free_T}
\end{equation}
where $\Delta g_{disorder}\left(T_0\right)$ is the free energy of a perfectly disordered random solution at a high temperature $T_0$, and $\langle \Delta e_f \rangle$ is the ensemble-averaged formation energy at intermediate temperature $\tau$, computed from finite-temperature simulations. The free energy computed in \cref{eq:free_T} accounts for the effects of short-range order (SRO) \cite{Wei2026aa}.

We estimated the ensemble-averaged energy in \cref{eq:free_T} using Monte Carlo (MC) simulations with the MLIP. Chemical decorations on simulation cells containing 1000 atoms were cooled from 10,000\textdegree C to 1500\textdegree C, using 1000\textdegree C decrements down to 3000\textdegree C followed by 100\textdegree C steps. At each temperature, 10,000 swaps were attempted with acceptance determined by the Metropolis criterion. Structures were relaxed to their zero-pressure state every 250 steps, and the relaxed configurations were used to compute ensemble-averaged formation energies. The temperature-dependent free energies were then obtained by thermodynamic integration (\cref{eq:free_T}). By using zero-pressure relaxed energies in the ensemble averages, this approach isolates the configurational entropy contribution to the free energy, analogous to on-lattice cluster expansion methods that have proven accurate for a wide range of multicomponent materials including metallic alloys, carbides, and nitrides.  

SRO was computed from the same atomic configurations used for \cref{eq:free_T} using Warren-Cowley parameters. The SRO parameter $\alpha_{ij}^{(d)}$ quantifies the extent of short-range order for each pair of metal species $i$ and $j$ at the $d$-th nearest neighbor distance. Negative values of $\alpha_{ij}^{(d)}$ indicate a tendency for attraction between elements $i$ and $j$, while positive values indicate a tendency for repulsion.

For disordered phases without SRO, corresponding to a perfectly random phase, the free energy can be approximated without finite-temperature simulations:
\begin{equation}
\Delta g_{\text{disorder}}(T) = \Delta h_{\text{disorder}} - T \Delta s_{\text{ideal}}
\label{eq:free_energy}
\end{equation}
where $\Delta h_{disorder} \approx \Delta e_f^{\text{disorder}}$ is the zero-temperature enthalpy of the perfectly disordered random solution, and $\Delta s_{\text{ideal}}$ is the ideal configurational entropy. The ideal solution entropy is estimated separately over the metal and carbon sublattices, where the metal sublattice accounts for all metallic elements present in the carbide and the carbon sublattice permits carbon-vacancy disorder. This free energy model is similar to that used in \cite{qureshi_predictive_2025}.

The enthalpy of the perfectly disordered random solution was computed using special quasirandom structures (SQS) \cite{zunger1990special} generated with the \texttt{CASM} \cite{thomas2013finite, puchala2013thermodynamics, van2018first} software package. Structures were constructed for equiatomic metal carbides with 1:1 metal-to-carbon ratio. The metal sublattice was populated by all possible equiatomic compositions containing 1–9 metals. To approximate random disorder, pair correlations were enforced up to the second nearest neighbor, while higher-order (three- and four-body) correlations were constrained at the first nearest neighbor.
We evaluated the stability of disordered alloys across all three structural prototypes, yielding 1533 unique compositions. All SQS structures were relaxed at 0\,K with the fine-tuned MACE potential before evaluating formation energies and elastic properties.

SQS structures provide a reasonable estimate of the disordered enthalpy, but the result can be sensitive to the degree of disorder enforced in the structure. As the range and many-body character of the relevant chemical interactions are not known \textit{a priori}, insufficient disorder in longer-range or multi-body correlations can bias the enthalpy estimate. To test this sensitivity, we also computed disordered enthalpies from chemical decorations generated through hybrid MC/MD simulations. These decorations sample the equilibrium state of disorder at elevated temperatures and therefore serve as an independent check on the SQS-derived values. Structures were initialized with random metal-site occupations on the rocksalt lattice, equilibrated at 3000\textdegree C for 1\,ps, then quenched to 1500\textdegree C in 100\textdegree C intervals, alternating 100 MC swaps with 100 molecular dynamics (MD) steps at each stage. After cooling, an additional 8000 MC/MD steps were performed at 1500\textdegree C to ensure convergence of local ordering. The final chemical decoration was then relaxed at 0\,K before computing the properties of the disordered phase.

The hardness of each carbide was estimated using established surrogate models. Mechanical properties such as hardness and fracture toughness depend on microstructural features and the ease of defect motion within the material \cite{watkins_insights_2024,WATKINS2025121350,Hang2017}.
While descriptors such as valence electron concentration (VEC) are often used to predict phase stability and mechanical properties, recent work has shown that VEC does not reliably capture elastic moduli or hardness in multicomponent TMCs \cite{Zhang:2024aa}. The MLIP developed in this study enables direct computation of elastic properties for complex compositions.
Several studies have shown that elastic properties, especially bulk and shear moduli, provide reliable proxies for estimating hardness \cite{sarker_high-entropy_2018,mazhnik_model_2019}. These models offer computational efficiency and are widely used for preliminary materials screening. 

Several models correlating elastic properties with hardness have been developed. Teter \cite{Teter1998} proposed a linear correlation between Vickers hardness $H_v$ and shear modulus $G$. Subsequent refinements by Chen \textit{et al.} \cite{CHEN20111275} and Tian \textit{et al.} \cite{TIAN201293} introduced nonlinear expressions that account for the influence of bulk modulus and the geometry of the Vickers indenter. More recently, Mazhnik and Oganov \cite{mazhnik_model_2019} developed an elasticity-based model that also accommodates auxetic materials, using Poisson's ratio $\nu$ and Young's modulus $E$ as inputs. A detailed description of the models is given in Supplementary \cref{SI-subsec:appendix:Hv}.

We estimated single-crystal elastic constants using the fine-tuned MACE potential. The full anisotropic elasticity tensor was then used to derive isotropic bulk modulus $B$, shear modulus $G$, and Young's modulus $E$ via the Voigt–Reuss–Hill averaging scheme. These values served as inputs to the Teter, Chen, Tian, and Mazhnik models to estimate the Vickers hardness for each carbide composition.

\section{Model Performance}
\label{sec:model_performance}

Accurate predictions of low-temperature phase stability are a prerequisite for using MLIPs to model the effects of crystallography and chemical order at elevated temperatures. Recent studies\cite{deng2024,D4FD00107A,Radova2025} have shown that foundation models can be fine-tuned to reproduce the subtle energy differences that govern low-temperature phase stability. To quantify the data efficiency of this approach, we trained a series of models on progressively larger fractions of the training set while keeping all other training parameters fixed. Subsets were drawn in 10\% increments (approximately 2600 structures) from 95\% of the full dataset, with the remaining 5\% held out for validation. All validation structures had formation energies below 0.5 eV/atom, ensuring that evaluations focused on thermodynamically relevant configurations. Each model was evaluated on this fixed validation set to assess accuracy as a function of training set size.

\begin{figure}[!ht]
  \centering
  \includegraphics[width=0.48\textwidth]{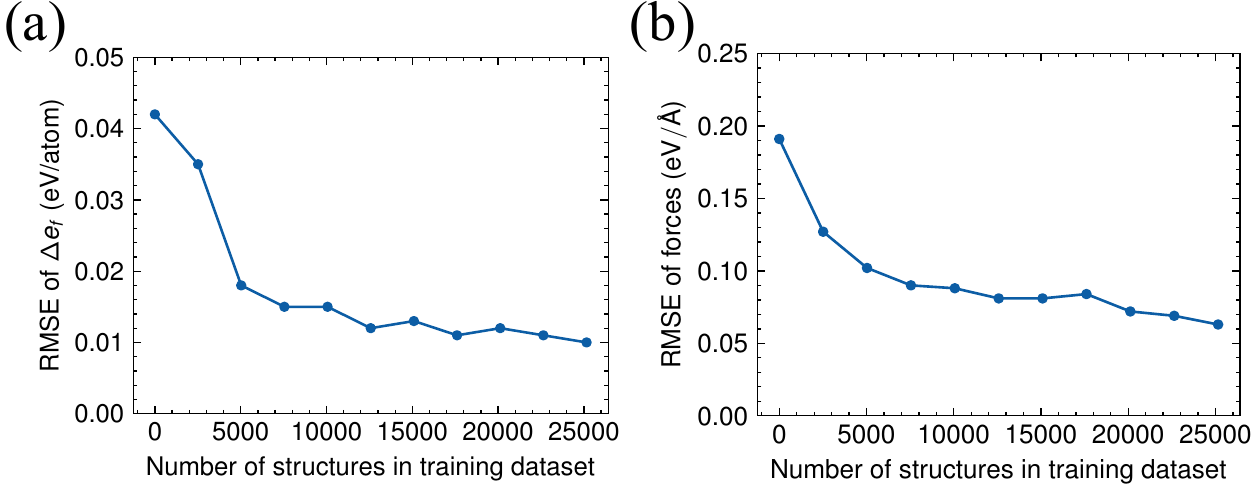}
  \caption{Root-mean-square errors (RMSEs) for formation energies and forces on a fixed validation set of low-energy structures (formation energy $<0.5$\,eV/atom) as a function of training set size. The error of the foundation MACE model corresponds to a training set of size zero. Errors plateau near 5,300 structures ($\sim$20\% of the training set).}
  \label{fig:learning_curve}
\end{figure}

\Cref{fig:learning_curve} shows the validation errors for formation energies and forces as a function of training set size. Both errors converge once approximately 20\% of the dataset (about 5300 structures) is included in training. Because validation errors are computed only over low-energy structures with formation energies below 0.5\,eV/atom, this convergence directly reflects the accuracy relevant to phase stability predictions. The rapid saturation of error metrics across the 10-component composition space of groups 4--6 carbides is consistent with prior observations of data-efficient fine-tuning of foundation MLIPs \cite{deng2024, D4FD00107A,Radova2025}.

Although the learning curves in \cref{fig:learning_curve} show that formation energy errors approach 10\,meV/atom with relatively little data, reproducing low-temperature phase stability can be more challenging\cite{piersante}.
To assess this, we computed convex hulls of formation energies for Mo carbides using three models: the \texttt{MACE-MPA-0} foundation model, a partially fine-tuned model trained on 20\% of the data (\texttt{MACE-20}), and a model fine-tuned on the entire training dataset (\texttt{MACE-100}).

\begin{figure}[!ht]
  \centering
  \includegraphics[width=0.48\textwidth]{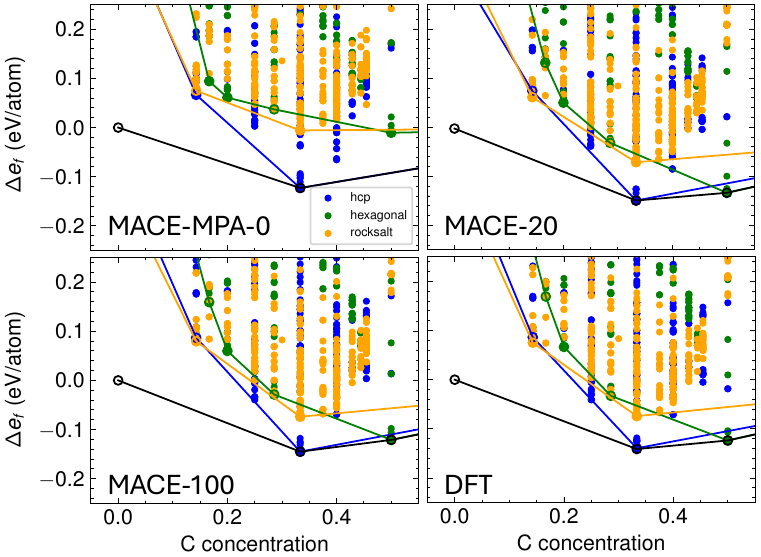}
  \caption{Convex hulls of formation energies for Mo-C at zero Kelvin, computed with three MACE models and DFT. Points and lines are colored by the structural prototype used to initialize each structure before relaxation: blue (hcp), green (hexagonal), orange (rocksalt). Open circles denote on-hull structures and the black line indicates the global convex hull.}
  \label{fig:hull_Mo}
\end{figure}

\Cref{fig:hull_Mo} compares the formation energies of carbon-vacancy orderings over the interstitial sites of molybdenum in the hcp, hexagonal and rocksalt crystal structures. DFT calculations predict two ground-state orderings in the Mo-C system. The equiatomic MoC contains carbon within the octahedral sites of a hexagonal metal host, while in Mo$_{2}$C, the carbon partially occupies the interstitial sites of molybdenum in the hcp crystal structure. The \texttt{MACE-MPA-0} foundation model fails to reproduce this behavior, incorrectly placing hexagonal MoC above the convex hull. Fine-tuning with just 20\% of the data (\texttt{MACE-20}) recovers a convex hull in excellent agreement with DFT. Training on the full dataset (\texttt{MACE-100}) yields only marginally better predictions, suggesting that foundation models fine-tuned on modest, targeted datasets accurately capture both global error metrics and phase stability.

\begin{figure}[!ht]
  \centering
  \includegraphics[width=0.48\textwidth]{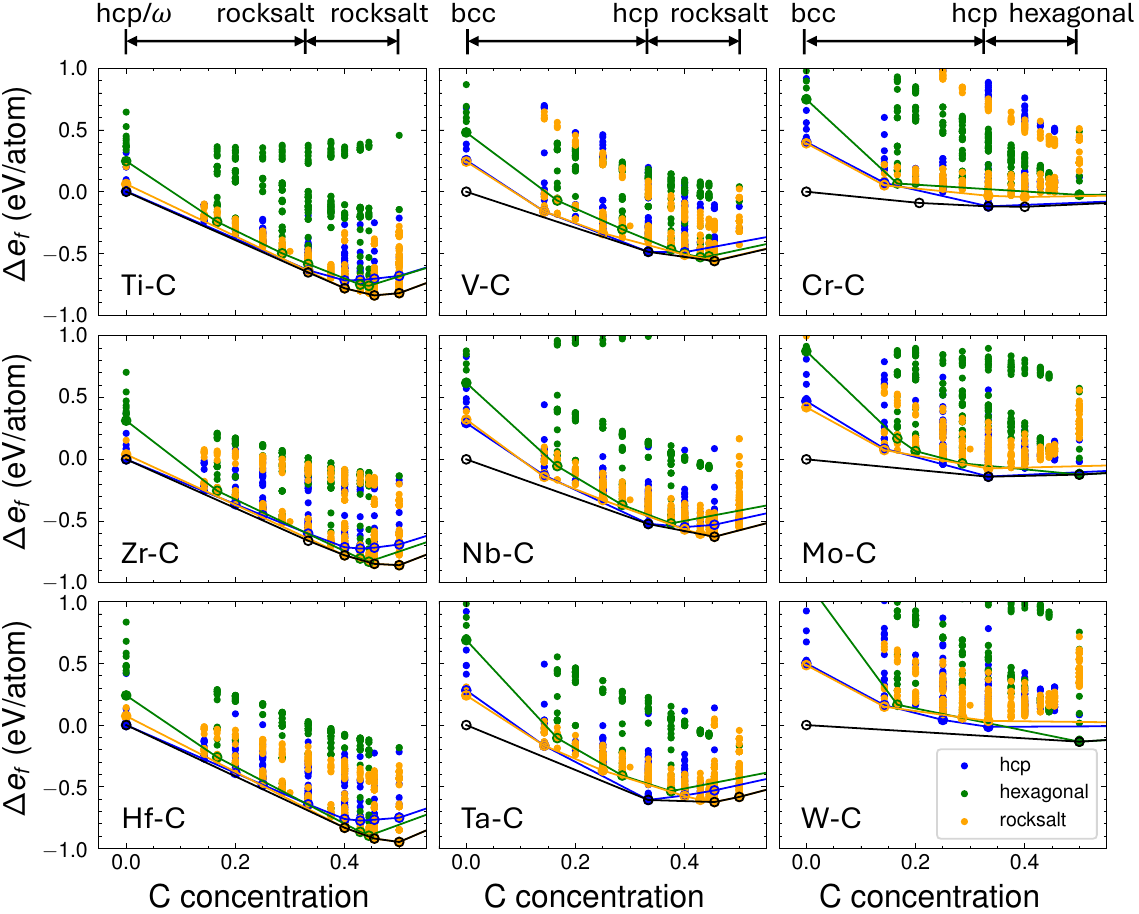}
  \caption{DFT-computed convex hulls for groups 4--6 transition metal carbides across three structural prototypes: hcp (blue), hexagonal (green), and rocksalt (orange). Each panel shows one metal-carbon binary. The black line denotes the global convex hull and open circles indicate stable structures.}
  \label{fig:formation_hull_DFT}
\end{figure}

\begin{figure}[!ht]
  \centering
  \includegraphics[width=0.48\textwidth]{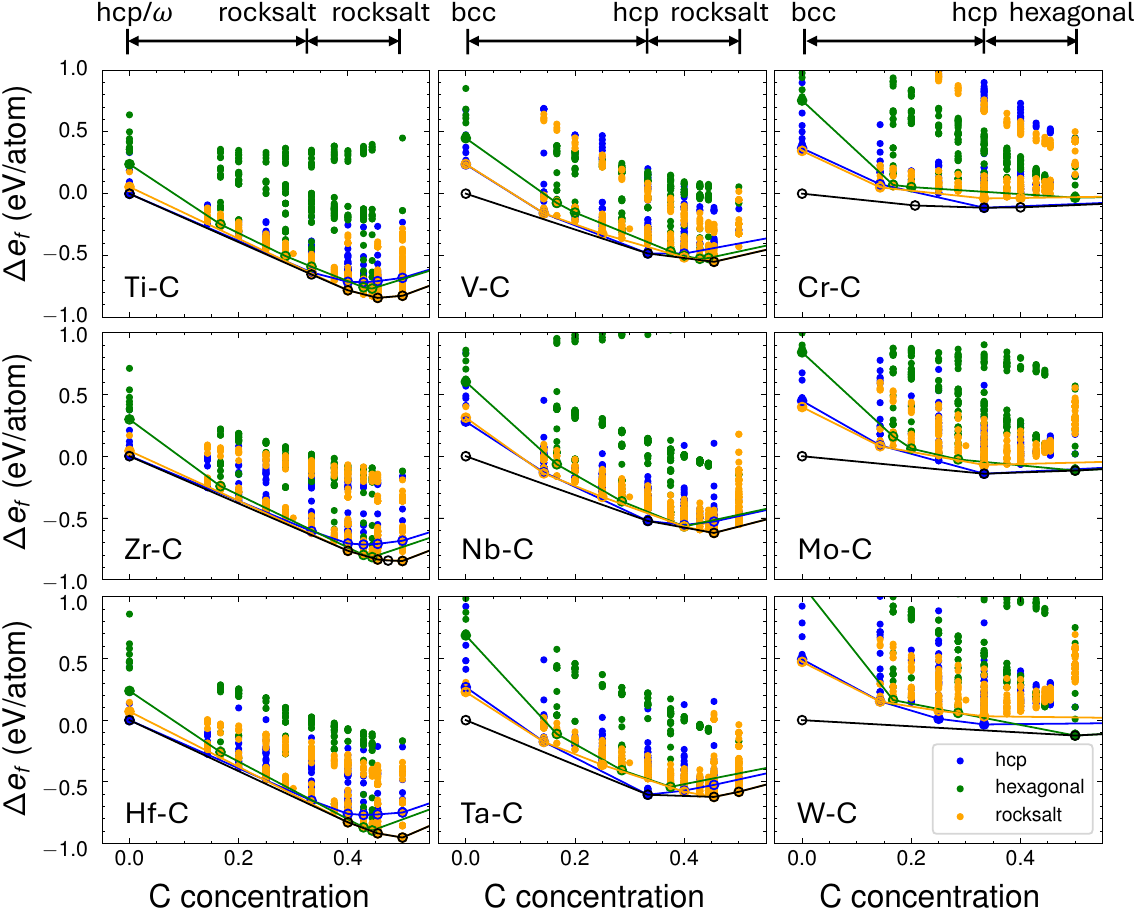}
  \caption{Convex hulls for groups 4--6 transition metal carbides predicted by the \texttt{MACE-100} model across three structural prototypes: hcp (blue), hexagonal (green), and rocksalt (orange). Each panel shows one metal-carbon binary. The black line denotes the global convex hull and open circles indicate stable structures.}
  \label{fig:formation_hull}
\end{figure}

\Cref{fig:formation_hull,fig:formation_hull_DFT} depict the DFT computed formation energies of carbon-vacancy orderings for all nine metal--carbon binaries composed of metals in groups 4--6 of the periodic table. Group 4 metals form rocksalt-based carbides across the full range of carbon composition, with the pure metal adopting either hcp or the $\omega$ crystal structure\cite{Gunda2018,NATARAJAN2020171}.
Group 5 metals crystallize in bcc in elemental form and form hcp-based carbides at low carbon compositions, transitioning to rocksalt at higher carbon content. Group 6 elements form low-energy hexagonal carbides at a carbon fraction $x_{C}=0.5$. At lower carbon stoichiometries, hcp-based carbides are stable for Cr and Mo, while W exhibits low-energy hcp-based carbides that remain thermodynamically metastable at 0\,K. DFT calculations indicate that the group 6 metals tend to form carbides with more complex crystal structures beyond the rocksalt, hcp, and hexagonal structures considered here. A comparison of the convex hulls predicted by the \texttt{MACE-100} model (\cref{fig:formation_hull}) with DFT (\cref{fig:formation_hull_DFT}) indicates excellent agreement between the fine-tuned potential and electronic structure calculations.

The \texttt{MACE-100} model achieves RMSEs of 92\,meV/atom (formation energy) and 135\,meV/\AA\ (forces) over the full dataset (Supplementary \cref{SI-fig:energy_force}). For low-energy structures with formation energies below 0.5\,eV/atom, these errors drop to 8\,meV/atom and 51\,meV/\AA, respectively. The model reproduces DFT-predicted hulls and correctly identifies ground-state structures across all nine single-metal systems, confirming its reliability for screening new compositions. Additional comparisons with \texttt{MACE-20} and \texttt{MACE-MPA-0} are provided in Supplementary \cref{SI-fig:formation_hull_mpa_0,SI-fig:formation_hull_mace20}. We use the \texttt{MACE-100} model to predict the properties of multicomponent carbides in the remainder of this study.

\section{Stability and hardness of multicomponent carbides}

The accuracy of the \texttt{MACE-100} model enables investigation of the finite-temperature thermodynamics and mechanical properties of multicomponent TMCs. We first benchmark our atomistic and free energy models against experimental reports of single-phase carbides and assess the impact of short-range order on phase stability. We then screen equiatomic compositions across the three structural prototypes (rocksalt, hcp, and hexagonal) to identify candidates that combine thermodynamic stability with high hardness.

\begin{figure}[!ht]
  \centering
  \includegraphics[width=0.48\textwidth]{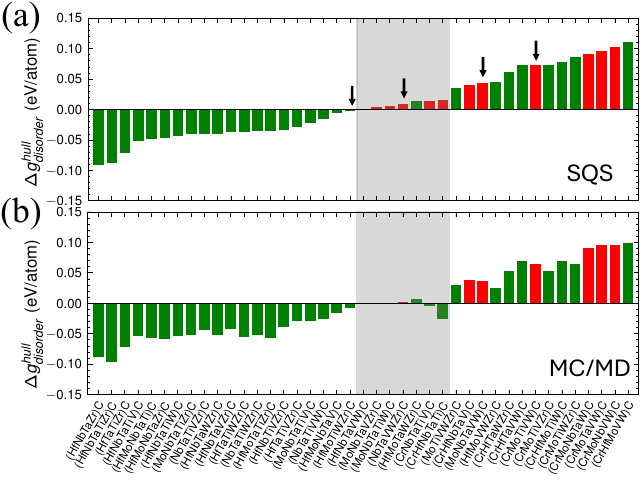}
  \caption{Synthesizability of equiatomic rocksalt TMCs at 1500\textdegree C predicted by \texttt{MACE-100}. Disordered phase enthalpies are computed from (a) SQS structures and (b) MC/MD simulations. Compositions with $\Delta g_{disorder}^{hull}<0$ are predicted to be single phase and those with $\Delta g_{disorder}^{hull}>0$ are predicted to be multiphase. Green markers indicate agreement with experimental reports \cite{divilov_disordered_2024, qureshi_predictive_2025, sarker_high-entropy_2018,hossain2021entropy, castle2018processing, HARRINGTON2019271, chicardi2019low, wei2019high, chicardi2020synthesis} and red markers indicate disagreement. Arrows mark compositions selected for SRO analysis. The grey region denotes compositions within 20\,meV/atom of the convex hull, where predictions may be sensitive to synthesis temperature and model error.
  }
  \label{fig:synthesizability}
\end{figure}

We benchmarked our model by comparing predicted synthesizability of multicomponent metal carbides against experimental data. Qureshi \textit{et al.} \cite{qureshi_predictive_2025} previously reported the synthesis of several five-metal carbide compositions as either single-phase or multiphase materials. Building on this work, we compiled a broader dataset of experimentally synthesized quaternary and quinary carbides from recent studies \cite{sarker_high-entropy_2018, hossain2021entropy, castle2018processing, HARRINGTON2019271, chicardi2019low, wei2019high, chicardi2020synthesis}. These carbides all crystallize in the rocksalt structure and were synthesized at temperatures between 1200\textdegree C and 2300\textdegree C. We evaluated the free energy of each composition at 1500\textdegree C and classified a disordered carbide as synthesizable if its free energy lies below the convex hull constructed from the 0\,K formation energies of the structures in our training dataset. The ordered phases are assumed to have negligible configurational entropy. \Cref{fig:synthesizability} reports the predicted hull distance $\Delta g_{disorder}^{hull}$ at 1500\textdegree C. Negative values indicate that the disordered phase is thermodynamically stable as a single phase, while positive values indicate that decomposition into competing phases is energetically favored.

\Cref{fig:synthesizability}a depicts $\Delta g_{disorder}^{hull}$ for 42 experimentally characterized compositions, computed with \cref{eq:free_energy} using the \texttt{MACE-100} model. Compositions are ranked by increasing hull distance, with mixing enthalpies estimated from SQS structures. Of these, 20 are predicted to be thermodynamically stable at 1500\textdegree C ($\Delta g_{disorder}^{hull} < 0$), and all 20 have been experimentally synthesized as single-phase disordered rocksalt carbides. Among the remaining 22 compositions, 10 are known to form multiphase microstructures, and our model correctly predicts their disordered phases to lie above the convex hull. The only discrepancies, shown in red in \cref{fig:synthesizability}a, arise from compositions where the model predicts multiphase behavior but experiments report single-phase synthesis.

The 12 incorrect predictions shown in red in \cref{fig:synthesizability}a fall into two groups. Nearly half correspond to compositions in the gray-shaded region with $\Delta g_{disorder}^{hull}$ between 0 and 20\,meV/atom, where the disordered phase is only marginally unstable. Predicted free energies for these compositions are sensitive to several factors including the assumed synthesis temperature, the neglect of short-range order in \cref{eq:free_energy}, the choice of SQS ordering, and errors in the MLIP. The average formation energy error of the \texttt{MACE-100} model for low-energy structures is approximately 8\,meV/atom, and free energy differences smaller than this value should be treated with caution.

\Cref{fig:synthesizability}b shows the same hull distances recomputed using mixing enthalpies from hybrid MC/MD simulations rather than SQS structures. For each composition, the MC/MD simulation generates a single chemical decoration of the metal sublattice representative of disorder at 1500\textdegree C, which is then relaxed at 0\,K to obtain the mixing enthalpy. The resulting free energies closely match those from SQS orderings, confirming that the choice of the precise chemical decoration to mimic the disordered phase has limited impact on the predictions.

Next we assess the impact of SRO on the free energy predictions of \cref{fig:synthesizability}. The free energy model in \cref{eq:free_energy} assumes perfect disorder, but at finite temperature SRO develops and reduces both enthalpy and entropy. SRO lowers the free energy of the disordered phase when the enthalpy reduction exceeds the corresponding entropy loss. To quantify this effect, we selected four representative compositions, indicated by arrows in \cref{fig:synthesizability}a, and computed their free energies with \cref{eq:free_T}. In all four cases, SRO lowers the free energy relative to the perfectly disordered reference, but only by a few meV/atom: 2.2\,meV/atom for (CrMoTiVW)C, 5.3\,meV/atom for (HfMoTiWZr)C, 0.8\,meV/atom for (MoNbTaVW)C, and 2.3\,meV/atom for (NbTaVWZr)C. These corrections are small enough to neglect for compositions well above the stability threshold (e.g., $\Delta g_{disorder}^{hull}>20$\,meV/atom). However, for compositions near the threshold ($\Delta g_{disorder}^{hull} \approx 0$), indicated by the shaded region in \cref{fig:synthesizability}a, SRO effects may shift the predicted free energy below the hull, potentially reversing the classification from multiphase to single-phase.

\begin{figure}[!ht]
\centering
\includegraphics[width=0.48\textwidth]{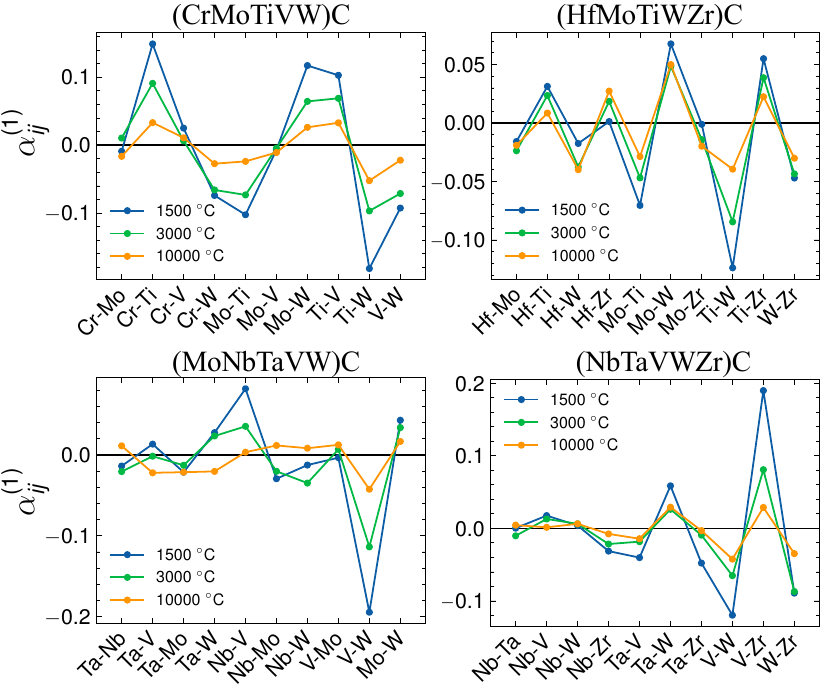}
\caption{First nearest-neighbor Warren-Cowley SRO parameters $\alpha_{ij}^{(1)}$ for (CrMoTiVW)C, (HfMoTiWZr)C, (MoNbTaVW)C, and (NbTaVWZr)C from MC simulations. 
Negative values indicate attraction between metal pairs and positive values indicate repulsion.}
\label{fig:sro}
\end{figure}

The finite-temperature simulations also yield the Warren-Cowley SRO parameters $\alpha_{ij}^{(1)}$ shown in \cref{fig:sro}. At $T=1500$\textdegree C, many metal pairs exhibit moderate short-range ordering, with $|\alpha_{ij}^{(1)}|>0.1$. For instance, Ti–W and V–W pairs exhibit relatively large negative $\alpha_{ij}^{(1)}$ values, indicating a tendency toward attraction, which can be attributed to their significant atomic size mismatch. In contrast, pairs of elements with similar atomic radii, such as Cr--Ti, tend to show positive $\alpha_{ij}^{(1)}$ values, reflecting a tendency toward repulsion.

SRO corrections and model error can account for some of the discrepancies discussed above, but six compositions in \cref{fig:synthesizability} have $\Delta g_{disorder}^{hull}$ values exceeding 30\,meV/atom. These outliers are unlikely to be explained by the assumptions or errors considered so far and are discussed in the next section.

\begin{table*}[!ht]\centering\footnotesize
\caption{Elastic properties and hardness. Bulk modulus $B$, shear modulus $G$, Young’s modulus $E$, and Vickers hardness $H_v$ from \texttt{MACE-100}, DFT, and experiments \cite{sarker_high-entropy_2018}. $H_v$ prediction uses Teter’s model.}
\label{tab:model_valid}
\resizebox{0.9\textwidth}{!}{
\begin{tabular}{ccccccccccccc}
\toprule
\multirow{2}{*}{Composition} & \multicolumn{3}{c}{$B$ (GPa)} & \multicolumn{3}{c}{$G$ (GPa)} & \multicolumn{3}{c}{$E$ (GPa)} & \multicolumn{3}{c}{$H_v$ (GPa)} \\  
\cmidrule(r){2-4} \cmidrule(r){5-7} \cmidrule(r){8-10} \cmidrule(r){11-13} 
& \texttt{MACE-100} & DFT & Exp. \cite{sarker_high-entropy_2018} 
& \texttt{MACE-100} & DFT & Exp. \cite{sarker_high-entropy_2018} 
& \texttt{MACE-100} & DFT & Exp. \cite{sarker_high-entropy_2018} 
& \texttt{MACE-100} & DFT & Exp. \cite{sarker_high-entropy_2018} \\ 
\midrule
TiC           & 249 & 249 & 255 & 190 & 180 & 207 & 455 & 436 & 489 & 29 & 27 & 31 \\
ZrC           & 218 & 221 & 216 & 156 & 162 & 169 & 378 & 391 & 402 & 24 & 25 & 24 \\
HfC           & 237 & 239 & 223 & 189 & 182 & 181 & 447 & 436 & 428 & 28 & 28 & 25 \\
VC            & 302 & 302 & 250 & 201 & 206 & 196 & 494 & 503 & 465 & 30 & 31 & 29 \\
NbC           & 302 & 299 & 246 & 203 & 202 & 177 & 497 & 495 & 429 & 31 & 30 & 17 \\
TaC           & 324 & 325 & 219 & 207 & 219 & 184 & 512 & 536 & 431 & 31 & 33 & 14 \\
(VNbMoTaW)C   & 317 & 322 & 278 & 173 & 190 & 226 & 439 & 476 & 533 & 26 & 29 & 27 \\
(TiVNbTaW)C   & 304 & 308 & 253 & 192 & 203 & 199 & 476 & 499 & 485 & 29 & 31 & 28 \\
(TiVNbHfTa)C  & 277 & 279 & 267 & 196 & 198 & 212 & 476 & 481 & 503 & 30 & 30 & 29 \\
(TiZrHfTaW)C  & 271 & 277 & 246 & 186 & 194 & 200 & 453 & 472 & 473 & 28 & 29 & 33 \\
(TiZrNbHfTa)C & 266 & 265 & 235 & 195 & 191 & 188 & 470 & 461 & 443 & 29 & 29 & 32 \\
(TiNbHfTaW)C  & 291 & 296 & 252 & 192 & 203 & 205 & 473 & 495 & 483 & 29 & 31 & 31 \\
\bottomrule
\end{tabular}
}
\end{table*}

Having established that the \texttt{MACE-100} model reliably predicts synthesizability, we now assess its accuracy for elastic properties and hardness. \Cref{tab:model_valid} compares bulk modulus, shear modulus, Young's modulus, and Vickers hardness from Teter's model for single-metal and multicomponent carbides, computed with \texttt{MACE-100} and DFT, against experimental data. MACE predictions agree well with DFT across all compositions, confirming the accuracy of the potential. Compared with experiment, hardness predictions are especially consistent for multicomponent carbides, indicating that Teter's model is a reliable surrogate for this composition class. For single-metal carbides, the model remains reasonable but overestimates hardness for TaC and NbC, where dislocation-mediated plasticity could play a significant role in determining hardness \cite{watkins_insights_2024,WATKINS2025121350}. Given our focus on multicomponent carbides, where Teter's model performs reliably, we adopt it as the primary screening tool.

\subsection{Integrating Stability and Hardness}
\label{sub:stab_hard}

The accuracy of the \texttt{MACE-100} model enables rapid screening of the nine-component chemical space of groups 4--6 transition metal carbides. We enumerated SQS-based chemical orderings for all carbides containing an equal amount of metals and carbon with the metal atoms arranged in the rocksalt, hexagonal, and hcp frameworks. Free energies for the disordered phase were evaluated with \cref{eq:free_energy} without accounting for SRO, while Vickers hardness was estimated with Teter's model.

\begin{figure}[!ht]
  \centering
  \includegraphics[width=0.4\textwidth]{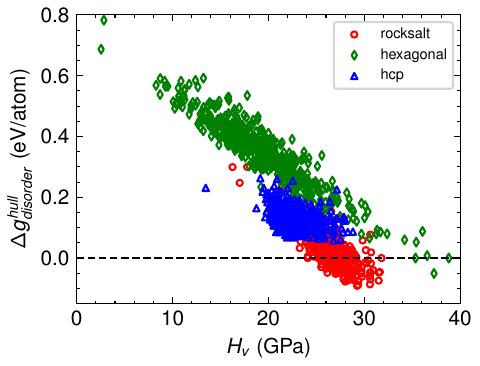}
  \caption{Predicted Vickers hardness (Teter model) versus free energy hull distance $\Delta g_{disorder}^{hull}$ at 1500\textdegree C for equiatomic carbides containing one to nine metals across the rocksalt, hcp, and hexagonal prototypes. Compositions below the dashed line ($\Delta g_{disorder}^{hull}=0$) are predicted to be thermodynamically stable as single-phase disordered carbides.}
  \label{fig:Hv_teter_ehull}
\end{figure}

\Cref{fig:Hv_teter_ehull} plots $\Delta g_{disorder}^{hull}$ against Vickers hardness for all equiatomic carbides containing one to nine metals across the three structural prototypes. The rocksalt structure accommodates the largest number of thermodynamically stable disordered compositions. Several hexagonal carbides are also predicted to be synthesizable at 1500\textdegree C, whereas no disordered hcp carbides are thermodynamically stable at this temperature. Among synthesizable compositions, rocksalt carbides reach hardness values exceeding 25\,GPa. Hexagonal carbides can achieve higher values, exceeding 32\,GPa, but only a few hexagonal compositions are predicted to be synthesizable.

\begin{figure}[!ht]
  \centering
  \includegraphics[width=0.47\textwidth]{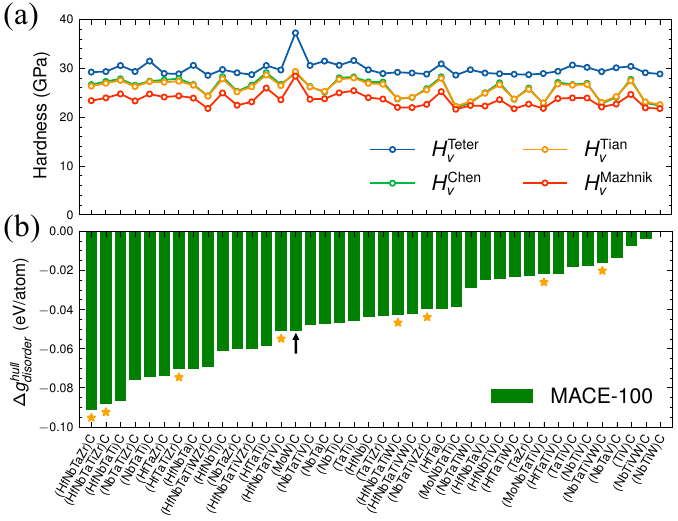}
  \caption{Forty equiatomic transition metal carbides with the highest predicted hardness that are also synthesizable at 1500\textdegree C. (a) Vickers hardness from four elasticity-based surrogate models and (b) distance above the convex hull of the disordered phase $\Delta g_{disorder}^{hull}$. Black arrows indicate compositions with the hexagonal prototype and all others adopt rocksalt. Orange stars mark experimentally validated predictions.}
  \label{fig:synthesizability_hv}
\end{figure}

The high-throughput screening of \cref{fig:Hv_teter_ehull} enables an exhaustive search of the nine-component chemical space for the hardest synthesizable multicomponent carbides. \Cref{fig:synthesizability_hv} presents the 40 multicomponent compositions with the highest predicted hardness, with $H_v$ evaluated using four elasticity-based models alongside $\Delta g_{disorder}^{hull}$ at 1500\textdegree C. The hardest multicomponent carbides are mixtures of three to five metals. Among rocksalt-based carbides, most contain elements from groups 4 and 5. The only hexagonal composition that is both stable and hard is (MoW)C, composed entirely of group 6 elements. Experimentally validated predictions are marked with orange stars in \cref{fig:synthesizability_hv}, further supporting the reliability of the screening framework. Many of the highest-hardness compositions identified here have not yet been synthesized, presenting concrete targets for experimental exploration.

\section{Stacking-ordered multicomponent carbides}
\label{sec:stacking}

The crystal structure of single-metal carbides depends on the group number of the metal. Group 4 and 5 metals form rocksalt carbides with ABC stacking of the metal layers, while group 6 metals form hexagonal carbides with AA stacking (\cref{fig:structure_prototypes}). Mixing metals from different groups could allow the stacking sequence to be tuned in multicomponent compositions, analogous to stacking-ordered phases in elemental metals and alloys. In Li, the fcc and hcp structures differ by only 1--2\,meV/atom at zero Kelvin, and an entire family of stacking-ordered structures (ABAC, ABCBC, 9R, 27R, etc.) are nearly degenerate in energy\cite{raju_natarajan_toward_2019}. Similar phases arise in multicomponent alloys that combine elements with different ground-state structures. In NiCoCr, mixing hcp Co, fcc Ni, and bcc Cr produces intermediate stacking sequences observed experimentally as stacking faults, and comparable behavior occurs in CuAg alloys \cite{raju_natarajan_toward_2019, BARUFFI2023115536, LI2024174091}.

The formation energies in \cref{fig:formation_hull_DFT} support this hypothesis, showing that increasing the fraction of group 6 metals destabilizes the rocksalt structure while favoring hexagonal stacking. For compositions containing both group 4/5 and group 6 metals, neither stacking sequence may represent the energetic minimum, and mixed-stacking configurations could be lower in energy.

To test whether stacking-ordered phases emerge in multicomponent TMCs, we first examined the effect of stacking sequence on formation energy in NbC, MoC, and their equiatomic solid solution (NbMo)C. Nb forms stable rocksalt carbides, while Mo favors hexagonal stacking (\cref{fig:formation_hull}). Following the method of Natarajan \textit{et al.} \cite{raju_natarajan_toward_2019}, we generated stacking variants by applying strain deformations to a bcc lattice and shuffling atomic layers. The $e_{2}$ strain order parameter, computed relative to a perfect bcc crystal, quantifies the degree of stacking transformation. Increasing $e_2$ from 0 to approximately 0.25 gradually converts the structure from hexagonal/hcp-like to rocksalt-like stacking. A detailed description of the order parameter and the crystallography of stacking-ordered phases can be found in \cite{raju_natarajan_toward_2019}. We populated the metal sublattice with Nb, Mo, or an equiatomic mixture of both, relaxed all structures with the \texttt{MACE-100} model, and evaluated their formation energies at zero Kelvin.

\begin{figure}[!ht]
\centering
\includegraphics[width=0.48\textwidth]{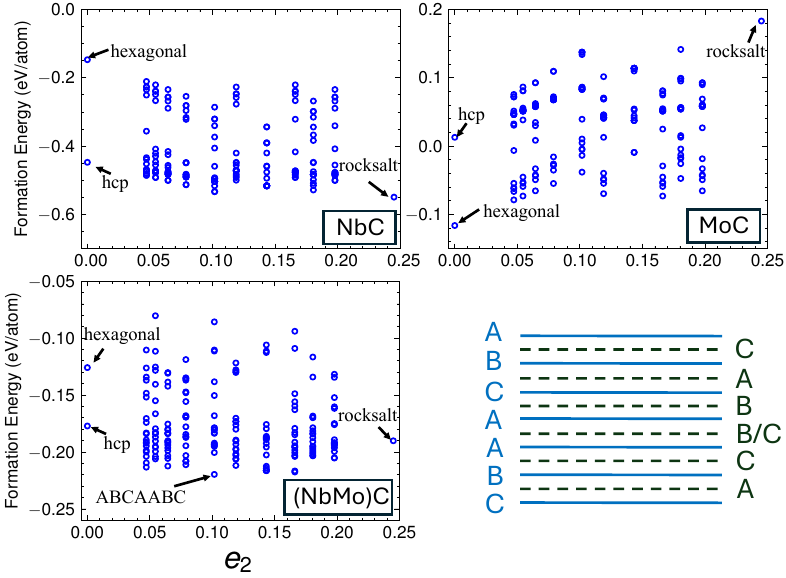}
\caption{Formation energies as a function of the stacking order parameter $e_2$ for NbC, MoC, and (NbMo)C. Increasing $e_2$ transforms the structure from hexagonal/hcp-like to rocksalt-like stacking. The lower right panel shows the ABCAABC stacking sequence of metal atoms (blue) and the corresponding carbon layers (black).}
\label{fig:stacking1}
\end{figure}

\Cref{fig:stacking1} shows the formation energies as a function of the $e_{2}$ order parameter. For NbC and MoC, the lowest-energy structures correspond to the simpler stacking sequences, rocksalt (fcc stacking) for NbC and hexagonal stacking for MoC, while deviations from these endpoints increase the energy. For the equiatomic (NbMo)C solid solution, however, intermediate stacking sequences yield lower formation energies than either endpoint. The ABCAABC stacking sequence is the most stable structure among those considered, nearly 35\,meV/atom lower in energy than the disordered rocksalt or hexagonal structure. 
Within this structure, Nb preferentially segregates to layers with fcc-like coordination, while Mo prefers hexagonal coordination, consistent with the stacking preferences of their respective single-metal carbides.

\begin{figure}[!ht]
\centering
\includegraphics[width=0.48\textwidth]{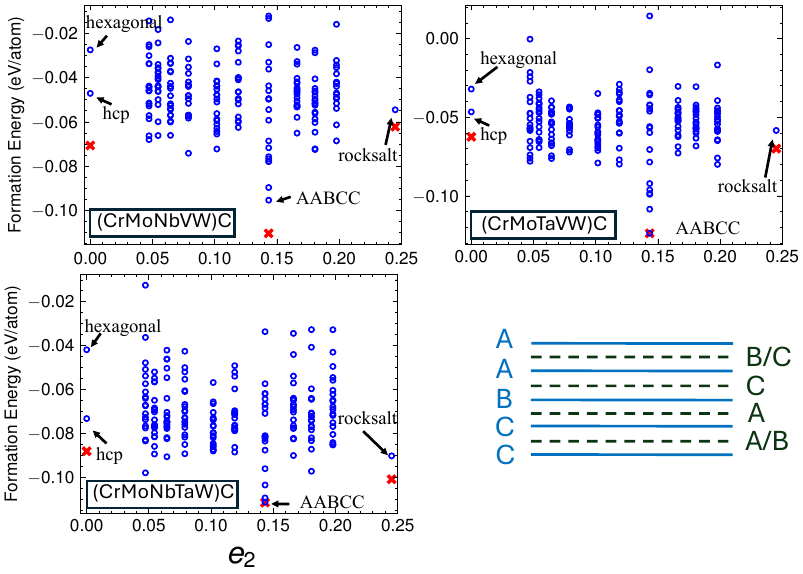}
\caption{Formation energies as a function of the stacking order parameter $e_2$ for three quinary carbides containing group 5 and group 6 metals. Red crosses mark DFT formation energies for the hcp, rocksalt, and AABCC stacking sequences. The lower right panel shows the AABCC stacking sequence of metal atoms (blue) and the corresponding carbon layers (black).}
\label{fig:stacking2}
\end{figure}

The binary (NbMo)C solid solution confirms that stacking-ordered phases can lower formation energies relative to the rocksalt and hexagonal end members. To test whether this behavior extends to more complex compositions, we enumerated stacking sequences for three quinary carbides, (CrMoNbVW)C, (CrMoTaVW)C, and (CrMoNbTaW)C, each containing a mixture of group 5 and group 6 metals. These compositions are predicted to be thermodynamically unstable as single-phase disordered rocksalt carbides (\cref{fig:synthesizability}a), yet single-phase disordered structures have been synthesized experimentally at all three compositions \cite{kaufmann_discovery_2020}. \Cref{fig:stacking2} shows the formation energies of stacking-ordered phases for these compositions. For all three, the AABCC stacking sequence yields formation energies well below those of the rocksalt structure, in some cases by nearly 50\,meV/atom. We validated these predictions with additional DFT calculations for structures with hcp, rocksalt, and AABCC stackings, shown as red crosses in \cref{fig:stacking2}. The DFT formation energies agree well with the \texttt{MACE-100} predictions, with differences within the fitting errors of the model, confirming that stacking-ordered phases are lower in energy than rocksalt or hexagonal structures for all three compositions.

The stacking-sequence analysis in \cref{fig:stacking2} used random chemical decorations of the metal sublattice. While random decorations adequately represent the energetics of simple stacking sequences such as rocksalt, hcp, and hexagonal, more complex sequences may favor chemical segregation across layers. To assess this effect, we performed Monte Carlo simulations on the AABCC stacking for (CrMoNbVW)C, (CrMoTaVW)C, and (CrMoNbTaW)C. The lowest-energy decorations were up to 40 meV/atom below the random ordering, confirming the importance of chemical segregation. Using these formation energies in the free energy model of \cref{eq:free_energy} yields $\Delta g_{disorder}^{hull}$ values of 23, 11, and 28\,meV/atom, respectively, much closer to the stability threshold ($\Delta g_{disorder}^{hull} \approx 0$) than those obtained from rocksalt structures (\cref{fig:synthesizability}). Synthesis temperatures above 1500\textdegree C or longer-period stacking sequences not explored in this study could further lower the free energy and bring these compositions within the range of synthesizability. 

Taken together, the results of \cref{fig:stacking1,fig:stacking2} reveal a new family of stacking-ordered phases in multicomponent carbides containing group 6 elements. The tendency of group 4/5 and group 6 metals to segregate into layers with fcc-like and hexagonal-like coordination implies that the ideal solution entropy used in \cref{eq:free_energy} may not accurately describe these phases, and more refined free energy models may be needed. The stacking variants explored here represent only a small subset of the combinatorially vast space of possible sequences. Nevertheless, adopting a non-rocksalt stacking sequence can lower the formation energy by several tens of meV/atom, sufficient in some cases to shift a composition from predicted instability to synthesizability.

\section{Discussion}
\label{sec:discussion}

\begin{figure*}[!ht]
  \centering
  \includegraphics[width=0.86\textwidth]{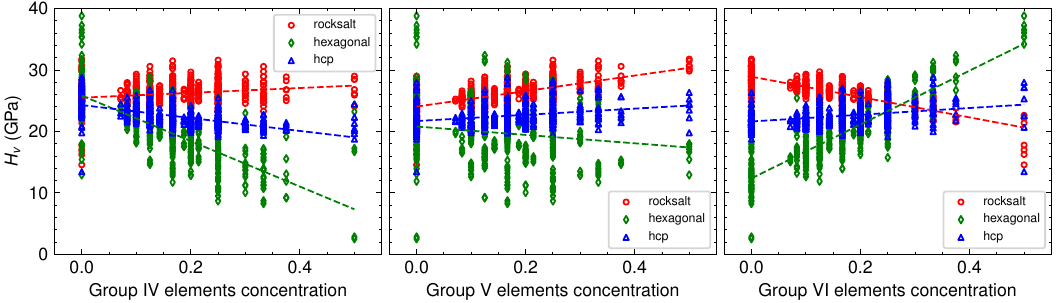}
  \caption{Vickers hardness (Teter model) for equiatomic transition metal carbides as a function of the fraction of metals from groups 4, 5, and 6, shown separately for the rocksalt, hcp, and hexagonal prototypes. Dashed lines are linear fits.}
  \label{fig:Hv_ele_group}
\end{figure*}

\begin{figure*}[!ht]
  \centering
  \includegraphics[width=0.86\textwidth]{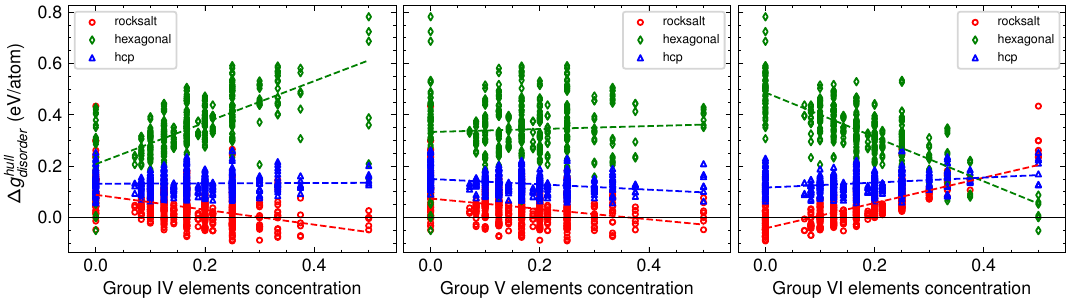}
  \caption{Distance above the convex hull of the disordered phase $\Delta g_{disorder}^{hull}$ for equiatomic transition metal carbides as a function of the fraction of metals from groups 4, 5, and 6, shown separately for the rocksalt, hcp, and hexagonal prototypes. Dashed lines are linear fits.}
  \label{fig:DeltaG_ele_group}
\end{figure*}

We screened over 1500 equiatomic carbide compositions across the rocksalt, hcp, and hexagonal crystal structures using a fine-tuned MACE interatomic potential, spanning the full nine-component space of groups 4--6 transition metals. Among the three prototypes, rocksalt accommodates the widest range of thermodynamically stable disordered carbides, particularly those rich in group 4 and 5 elements, whereas only a limited number of hexagonal compositions meet both stability and hardness criteria. Fine-tuning the MACE foundation model on only ${\sim}20\%$ (${\sim}5300$ structures) of the full DFT dataset yields near-DFT accuracy, enabling this high-throughput exploration at a fraction of the computational cost.

The free energy model used for screening assumes perfectly random mixing on the metal sublattice (\cref{eq:free_energy}). Our Monte Carlo simulations show that short-range order corrections are small, typically a few meV/atom, and that SRO-adjusted free energies closely track those computed from SQS supercells. Hybrid MC/MD simulations similarly yield formation energies close to those of the corresponding random configurations. These results indicate that treating the disordered phase as a perfectly random solid solution is a practical approximation for high-throughput screening. Accounting for SRO becomes important for compositions close to the synthesizability threshold, where free energy differences are on the order of a few meV/atom and SRO corrections may shift the predicted stability. For such compositions, more refined free energy models are necessary.

Beyond screening for synthesizable carbides with favorable mechanical properties, this work reveals a new family of stacking-ordered phases in multicomponent carbides containing group 6 metals. Intermediate stacking arrangements of the metal sublattice between the rocksalt (ABC) and hexagonal (AA) end members can lower formation energies significantly, sufficient in some cases to shift a composition from predicted instability to synthesizability. These phases arise because group 4/5 and group 6 metals preferentially segregate into layers with fcc-like and hexagonal-like coordination, respectively, producing mixed stacking sequences that neither end-member structure can access. This tendency implies that the ideal solution entropy assumed in \cref{eq:free_energy} may not accurately describe stacking-ordered phases, and more refined free energy treatments are required. The stacking variants explored here represent only a small subset of the combinatorially vast configuration space, and an exhaustive search over stacking sequences and metal-atom distributions remains a target for future work. Existing experiments on multicomponent carbides have focused exclusively on disordered rocksalt structures, and our results, supported by DFT calculations, suggest that stacking-ordered phases should be experimentally accessible and merit targeted synthesis.

The group number of the constituent metals governs both hardness and phase stability across all three structural prototypes as illustrated in \cref{fig:Hv_ele_group,fig:DeltaG_ele_group}. Increasing the fraction of group 4 and 5 elements stabilizes disordered rocksalt carbides and increases their hardness, consistent with the preference of these elements for rocksalt-based crystal structures. \Cref{fig:DeltaG_ele_group} shows that increasing the fraction of group 6 elements has the opposite effect on stability, destabilizing the rocksalt phase while favoring hexagonal structures. The corresponding hardness trends in \cref{fig:Hv_ele_group} reveal that group 6 elements simultaneously raise hardness in hexagonal carbides. These opposing trends explain why rocksalt multicomponent carbides are overwhelmingly composed of group 4/5 metals, while the few viable hexagonal compositions are enriched in group 6 elements. Although disordered hexagonal carbides can be considerably harder than their rocksalt or hcp counterparts as shown by \cref{fig:Hv_ele_group}, stabilizing them through multi-metal mixing remains difficult because adding group 4/5 metals rapidly favors the competing rocksalt phase. These trends emerge directly from MLIP-based property predictions, without relying on empirical chemical surrogates such as valence electron count or local lattice distortion.

\begin{figure}[!ht]
  \centering
  \includegraphics[width=0.4\textwidth]{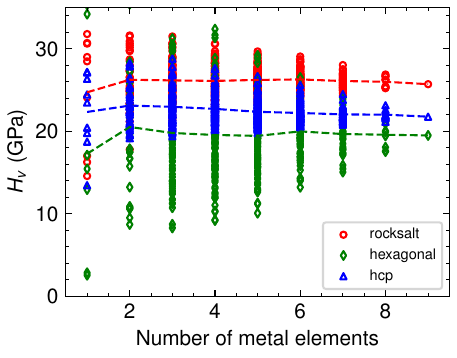}
  \caption{Predicted Vickers hardness $H_v$ (Teter model) as a function of the number of constituent metals for equiatomic transition metal carbides across the rocksalt, hcp, and hexagonal prototypes. Dashed lines indicate the average hardness for each metal count.}
  \label{fig:Hv_teter_metal_n}
\end{figure}

The MLIP also allows us to assess the effect of the number of elements on hardness. \Cref{fig:Hv_teter_metal_n} shows that increasing the number of metal components does not systematically increase hardness across any of the three crystal structures. Several binary and ternary compositions achieve $H_v$ values comparable to or exceeding those of quinary systems, confirming that the group identity of the constituent metals governs hardness more than compositional complexity alone. These trends are independent of the choice of hardness surrogate model, as the Chen, Tian, and Mazhnik models all reproduce the same qualitative behavior (Supplementary \cref{SI-fig:Hv_models,SI-fig:Hv_metal_n_models}).

The hardness trends in \cref{fig:Hv_teter_metal_n,fig:Hv_ele_group} rely on elastic surrogate models, which do not capture the complex deformation mechanisms activated during indentation. Among the four models evaluated, Teter's model shows the strongest agreement with DFT and experimental data and is adopted as the primary screening tool. However, plastic flow under a sharp indenter is governed by dislocation nucleation and motion, which depend on the active slip systems, unstable stacking-fault energies $\gamma_{\mathrm{us}}$, Peierls stresses, and barriers for cross-slip or kink nucleation \cite{watkins_insights_2024}. Elastic moduli do not capture these quantities, and the predictions may therefore miss disorder-induced strengthening or softening effects observed in metallic high-entropy alloys \cite{VARVENNE2016164,VARVENNE2017660,LIU2024119471,MARESCA2020144,MARESCA2020235}. Real TMCs also contain carbon vacancies, anti-site defects, and oxygen impurities that alter dislocation behavior, and because this work considers only equiatomic carbides, such effects are not reflected in the screening. For instance, sub-equiatomic TaC$_{x}$ ($x<1$) exhibits higher hardness correlated with the dominant slip system shifting from $\{111\}$ to $\{110\}$ \cite{watkins_insights_2024}. Elastic models could be augmented with plasticity-aware descriptors such as unstable stacking-fault energies, ideal shear strengths, or Peierls stress estimates derived from $\gamma$ surfaces and dislocation core widths \cite{Karumuri2025}. Machine-learned potentials such as MACE could compute these descriptors efficiently, offering a path to more physically grounded hardness predictions. More broadly, the interplay between stacking order, carbon-sublattice disorder in substoichiometric carbides, and defect-mediated plasticity offers a rich design space that connects synthesizability to mechanical performance beyond what elastic models alone can capture. Nevertheless, elasticity-based surrogate models remain practical for high-throughput screening of high-dimensional composition spaces and could serve as a first-pass tool before applying more detailed plasticity-based models to a smaller set of promising candidates.

\section{Conclusion}
\label{sec:conclusion}

This study combines a fine-tuned MACE interatomic potential with free energy models and elasticity-based hardness surrogates to screen over 1500 equiatomic compositions across rocksalt, hexagonal, and hcp prototypes, spanning the full nine-component chemical space of groups 4--6 transition metal carbides. Synthesizability predictions at 1500\textdegree C agree well with experimental reports, and short-range order corrections are small enough that an ideal solution approximation suffices for high-throughput screening. The group number of the constituent metals correlates strongly with both thermodynamic stability and hardness, with group 4 and 5 elements favoring stable rocksalt carbides and group 6 elements enhancing hardness in the hexagonal structure. For compositions mixing elements from both groups, we identify a new family of stacking-ordered phases with formation energies significantly below those of the simpler end-member structures. Experimental validation of the predicted stacking-ordered phases is an important next step. More broadly, the combination of fine-tuned machine-learned potentials with statistical mechanics methods provides a scalable framework for exploring the vast compositional and structural design space of multicomponent ceramics, extending well beyond the carbide chemistries considered here.

\begin{acknowledgments}
  ARN is grateful to support from EUROfusion through the Enabling Research programme (grant number EnR-04-EPFL-01).
\end{acknowledgments}

\bibliography{references}

\begin{thebibliography}{0}%
\makeatletter
\providecommand \@ifxundefined [1]{%
 \@ifx{#1\undefined}
}%
\providecommand \@ifnum [1]{%
 \ifnum #1\expandafter \@firstoftwo
 \else \expandafter \@secondoftwo
 \fi
}%
\providecommand \@ifx [1]{%
 \ifx #1\expandafter \@firstoftwo
 \else \expandafter \@secondoftwo
 \fi
}%
\providecommand \natexlab [1]{#1}%
\providecommand \enquote  [1]{``#1''}%
\providecommand \bibnamefont  [1]{#1}%
\providecommand \bibfnamefont [1]{#1}%
\providecommand \citenamefont [1]{#1}%
\providecommand \href@noop [0]{\@secondoftwo}%
\providecommand \href [0]{\begingroup \@sanitize@url \@href}%
\providecommand \@href[1]{\@@startlink{#1}\@@href}%
\providecommand \@@href[1]{\endgroup#1\@@endlink}%
\providecommand \@sanitize@url [0]{\catcode `\\12\catcode `\$12\catcode `\&12\catcode `\#12\catcode `\^12\catcode `\_12\catcode `\%12\relax}%
\providecommand \@@startlink[1]{}%
\providecommand \@@endlink[0]{}%
\providecommand \url  [0]{\begingroup\@sanitize@url \@url }%
\providecommand \@url [1]{\endgroup\@href {#1}{\urlprefix }}%
\providecommand \urlprefix  [0]{URL }%
\providecommand \Eprint [0]{\href }%
\providecommand \doibase [0]{https://doi.org/}%
\providecommand \selectlanguage [0]{\@gobble}%
\providecommand \bibinfo  [0]{\@secondoftwo}%
\providecommand \bibfield  [0]{\@secondoftwo}%
\providecommand \translation [1]{[#1]}%
\providecommand \BibitemOpen [0]{}%
\providecommand \bibitemStop [0]{}%
\providecommand \bibitemNoStop [0]{.\EOS\space}%
\providecommand \EOS [0]{\spacefactor3000\relax}%
\providecommand \BibitemShut  [1]{\csname bibitem#1\endcsname}%
\let\auto@bib@innerbib\@empty
\end{thebibliography}%


\begin{thebibliography}{56}%
\makeatletter
\providecommand \@ifxundefined [1]{%
 \@ifx{#1\undefined}
}%
\providecommand \@ifnum [1]{%
 \ifnum #1\expandafter \@firstoftwo
 \else \expandafter \@secondoftwo
 \fi
}%
\providecommand \@ifx [1]{%
 \ifx #1\expandafter \@firstoftwo
 \else \expandafter \@secondoftwo
 \fi
}%
\providecommand \natexlab [1]{#1}%
\providecommand \enquote  [1]{``#1''}%
\providecommand \bibnamefont  [1]{#1}%
\providecommand \bibfnamefont [1]{#1}%
\providecommand \citenamefont [1]{#1}%
\providecommand \href@noop [0]{\@secondoftwo}%
\providecommand \href [0]{\begingroup \@sanitize@url \@href}%
\providecommand \@href[1]{\@@startlink{#1}\@@href}%
\providecommand \@@href[1]{\endgroup#1\@@endlink}%
\providecommand \@sanitize@url [0]{\catcode `\\12\catcode `\$12\catcode `\&12\catcode `\#12\catcode `\^12\catcode `\_12\catcode `\%12\relax}%
\providecommand \@@startlink[1]{}%
\providecommand \@@endlink[0]{}%
\providecommand \url  [0]{\begingroup\@sanitize@url \@url }%
\providecommand \@url [1]{\endgroup\@href {#1}{\urlprefix }}%
\providecommand \urlprefix  [0]{URL }%
\providecommand \Eprint [0]{\href }%
\providecommand \doibase [0]{https://doi.org/}%
\providecommand \selectlanguage [0]{\@gobble}%
\providecommand \bibinfo  [0]{\@secondoftwo}%
\providecommand \bibfield  [0]{\@secondoftwo}%
\providecommand \translation [1]{[#1]}%
\providecommand \BibitemOpen [0]{}%
\providecommand \bibitemStop [0]{}%
\providecommand \bibitemNoStop [0]{.\EOS\space}%
\providecommand \EOS [0]{\spacefactor3000\relax}%
\providecommand \BibitemShut  [1]{\csname bibitem#1\endcsname}%
\let\auto@bib@innerbib\@empty
\bibitem [{\citenamefont {Wyatt}\ \emph {et~al.}(2024)\citenamefont {Wyatt}, \citenamefont {Nemani}, \citenamefont {Hilmas}, \citenamefont {Opila},\ and\ \citenamefont {Anasori}}]{Wyatt:2024aa}%
  \BibitemOpen
  \bibfield  {author} {\bibinfo {author} {\bibfnamefont {B.~C.}\ \bibnamefont {Wyatt}}, \bibinfo {author} {\bibfnamefont {S.~K.}\ \bibnamefont {Nemani}}, \bibinfo {author} {\bibfnamefont {G.~E.}\ \bibnamefont {Hilmas}}, \bibinfo {author} {\bibfnamefont {E.~J.}\ \bibnamefont {Opila}},\ and\ \bibinfo {author} {\bibfnamefont {B.}~\bibnamefont {Anasori}},\ }\bibfield  {title} {\bibinfo {title} {Ultra-high temperature ceramics for extreme environments},\ }\href@noop {} {\bibfield  {journal} {\bibinfo  {journal} {Nature Reviews Materials}\ }\textbf {\bibinfo {volume} {9}},\ \bibinfo {pages} {773} (\bibinfo {year} {2024})}\BibitemShut {NoStop}%
\bibitem [{\citenamefont {Williams}(1971)}]{WILLIAMS197157}%
  \BibitemOpen
  \bibfield  {author} {\bibinfo {author} {\bibfnamefont {W.~S.}\ \bibnamefont {Williams}},\ }\bibfield  {title} {\bibinfo {title} {Transition-metal carbides},\ }\href@noop {} {\bibfield  {journal} {\bibinfo  {journal} {Progress in Solid State Chemistry}\ }\textbf {\bibinfo {volume} {6}},\ \bibinfo {pages} {57} (\bibinfo {year} {1971})}\BibitemShut {NoStop}%
\bibitem [{\citenamefont {Wang}\ \emph {et~al.}(2020)\citenamefont {Wang}, \citenamefont {Zhang}, \citenamefont {Yan}, \citenamefont {Lu}, \citenamefont {Nastasi}, \citenamefont {Chen},\ and\ \citenamefont {Cui}}]{Wang2020}%
  \BibitemOpen
  \bibfield  {author} {\bibinfo {author} {\bibfnamefont {F.}~\bibnamefont {Wang}}, \bibinfo {author} {\bibfnamefont {X.}~\bibnamefont {Zhang}}, \bibinfo {author} {\bibfnamefont {X.}~\bibnamefont {Yan}}, \bibinfo {author} {\bibfnamefont {Y.}~\bibnamefont {Lu}}, \bibinfo {author} {\bibfnamefont {M.}~\bibnamefont {Nastasi}}, \bibinfo {author} {\bibfnamefont {Y.}~\bibnamefont {Chen}},\ and\ \bibinfo {author} {\bibfnamefont {B.}~\bibnamefont {Cui}},\ }\bibfield  {title} {\bibinfo {title} {The effect of submicron grain size on thermal stability and mechanical properties of high-entropy carbide ceramics},\ }\href@noop {} {\bibfield  {journal} {\bibinfo  {journal} {Journal of the American Ceramic Society}\ }\textbf {\bibinfo {volume} {103}},\ \bibinfo {pages} {4463} (\bibinfo {year} {2020})}\BibitemShut {NoStop}%
\bibitem [{\citenamefont {Zhang}\ and\ \citenamefont {Reece}(2019)}]{zhang_review_2019}%
  \BibitemOpen
  \bibfield  {author} {\bibinfo {author} {\bibfnamefont {R.-Z.}\ \bibnamefont {Zhang}}\ and\ \bibinfo {author} {\bibfnamefont {M.~J.}\ \bibnamefont {Reece}},\ }\bibfield  {title} {\bibinfo {title} {Review of high entropy ceramics: design, synthesis, structure and properties},\ }\href@noop {} {\bibfield  {journal} {\bibinfo  {journal} {Journal of Materials Chemistry A}\ }\textbf {\bibinfo {volume} {7}},\ \bibinfo {pages} {22148} (\bibinfo {year} {2019})}\BibitemShut {NoStop}%
\bibitem [{\citenamefont {Prats}\ and\ \citenamefont {Stamatakis}(2024)}]{prats_transition_2024}%
  \BibitemOpen
  \bibfield  {author} {\bibinfo {author} {\bibfnamefont {H.}~\bibnamefont {Prats}}\ and\ \bibinfo {author} {\bibfnamefont {M.}~\bibnamefont {Stamatakis}},\ }\bibfield  {title} {\bibinfo {title} {Transition {Metal} {Carbides} as {Supports} for {Catalytic} {Metal} {Particles}: {Recent} {Progress} and {Opportunities}},\ }\href@noop {} {\bibfield  {journal} {\bibinfo  {journal} {The Journal of Physical Chemistry Letters}\ }\textbf {\bibinfo {volume} {15}},\ \bibinfo {pages} {3450} (\bibinfo {year} {2024})}\BibitemShut {NoStop}%
\bibitem [{\citenamefont {Bie}\ \emph {et~al.}(2026)\citenamefont {Bie}, \citenamefont {Ren}, \citenamefont {Zhou}, \citenamefont {Brahimi}, \citenamefont {Yue},\ and\ \citenamefont {Song}}]{BIE2026153140}%
  \BibitemOpen
  \bibfield  {author} {\bibinfo {author} {\bibfnamefont {X.}~\bibnamefont {Bie}}, \bibinfo {author} {\bibfnamefont {B.}~\bibnamefont {Ren}}, \bibinfo {author} {\bibfnamefont {X.}~\bibnamefont {Zhou}}, \bibinfo {author} {\bibfnamefont {S.}~\bibnamefont {Brahimi}}, \bibinfo {author} {\bibfnamefont {S.}~\bibnamefont {Yue}},\ and\ \bibinfo {author} {\bibfnamefont {J.}~\bibnamefont {Song}},\ }\bibfield  {title} {\bibinfo {title} {Hydrogen trapping and diffusion in sub-stoichiometric vanadium and niobium carbide precipitates in high-strength steels},\ }\href@noop {} {\bibfield  {journal} {\bibinfo  {journal} {International Journal of Hydrogen Energy}\ }\textbf {\bibinfo {volume} {203}},\ \bibinfo {pages} {153140} (\bibinfo {year} {2026})}\BibitemShut {NoStop}%
\bibitem [{\citenamefont {Gunda}\ and\ \citenamefont {Van~der Ven}(2018)}]{Gunda2018}%
  \BibitemOpen
  \bibfield  {author} {\bibinfo {author} {\bibfnamefont {N.~S.~H.}\ \bibnamefont {Gunda}}\ and\ \bibinfo {author} {\bibfnamefont {A.}~\bibnamefont {Van~der Ven}},\ }\bibfield  {title} {\bibinfo {title} {First-principles insights on phase stability of titanium interstitial alloys},\ }\href@noop {} {\bibfield  {journal} {\bibinfo  {journal} {Phys. Rev. Mater.}\ }\textbf {\bibinfo {volume} {2}},\ \bibinfo {pages} {083602} (\bibinfo {year} {2018})}\BibitemShut {NoStop}%
\bibitem [{\citenamefont {Sarker}\ \emph {et~al.}(2018)\citenamefont {Sarker}, \citenamefont {Harrington}, \citenamefont {Toher}, \citenamefont {Oses}, \citenamefont {Samiee}, \citenamefont {Maria}, \citenamefont {Brenner}, \citenamefont {Vecchio},\ and\ \citenamefont {Curtarolo}}]{sarker_high-entropy_2018}%
  \BibitemOpen
  \bibfield  {author} {\bibinfo {author} {\bibfnamefont {P.}~\bibnamefont {Sarker}}, \bibinfo {author} {\bibfnamefont {T.}~\bibnamefont {Harrington}}, \bibinfo {author} {\bibfnamefont {C.}~\bibnamefont {Toher}}, \bibinfo {author} {\bibfnamefont {C.}~\bibnamefont {Oses}}, \bibinfo {author} {\bibfnamefont {M.}~\bibnamefont {Samiee}}, \bibinfo {author} {\bibfnamefont {J.-P.}\ \bibnamefont {Maria}}, \bibinfo {author} {\bibfnamefont {D.~W.}\ \bibnamefont {Brenner}}, \bibinfo {author} {\bibfnamefont {K.~S.}\ \bibnamefont {Vecchio}},\ and\ \bibinfo {author} {\bibfnamefont {S.}~\bibnamefont {Curtarolo}},\ }\bibfield  {title} {\bibinfo {title} {High-entropy high-hardness metal carbides discovered by entropy descriptors},\ }\href@noop {} {\bibfield  {journal} {\bibinfo  {journal} {Nature Communications}\ }\textbf {\bibinfo {volume} {9}},\ \bibinfo {pages} {4980} (\bibinfo {year} {2018})}\BibitemShut {NoStop}%
\bibitem [{\citenamefont {Dey}\ \emph {et~al.}(2024)\citenamefont {Dey}, \citenamefont {Liang},\ and\ \citenamefont {Yu}}]{dey_mixed_2024}%
  \BibitemOpen
  \bibfield  {author} {\bibinfo {author} {\bibfnamefont {D.}~\bibnamefont {Dey}}, \bibinfo {author} {\bibfnamefont {L.}~\bibnamefont {Liang}},\ and\ \bibinfo {author} {\bibfnamefont {L.}~\bibnamefont {Yu}},\ }\bibfield  {title} {\bibinfo {title} {Mixed {Enthalpy}--{Entropy} {Descriptor} for the {Rational} {Design} of {Synthesizable} {High}-{Entropy} {Materials} {Over} {Vast} {Chemical} {Spaces}},\ }\href@noop {} {\bibfield  {journal} {\bibinfo  {journal} {Journal of the American Chemical Society}\ }\textbf {\bibinfo {volume} {146}},\ \bibinfo {pages} {5142} (\bibinfo {year} {2024})}\BibitemShut {NoStop}%
\bibitem [{\citenamefont {Divilov}\ \emph {et~al.}(2024)\citenamefont {Divilov}, \citenamefont {Eckert}, \citenamefont {Hicks}, \citenamefont {Oses}, \citenamefont {Toher}, \citenamefont {Friedrich}, \citenamefont {Esters}, \citenamefont {Mehl}, \citenamefont {Zettel}, \citenamefont {Lederer}, \citenamefont {Zurek}, \citenamefont {Maria}, \citenamefont {Brenner}, \citenamefont {Campilongo}, \citenamefont {Filipovi{\'c}}, \citenamefont {Fahrenholtz}, \citenamefont {Ryan}, \citenamefont {DeSalle}, \citenamefont {Crealese}, \citenamefont {Wolfe}, \citenamefont {Calzolari},\ and\ \citenamefont {Curtarolo}}]{divilov_disordered_2024}%
  \BibitemOpen
  \bibfield  {author} {\bibinfo {author} {\bibfnamefont {S.}~\bibnamefont {Divilov}}, \bibinfo {author} {\bibfnamefont {H.}~\bibnamefont {Eckert}}, \bibinfo {author} {\bibfnamefont {D.}~\bibnamefont {Hicks}}, \bibinfo {author} {\bibfnamefont {C.}~\bibnamefont {Oses}}, \bibinfo {author} {\bibfnamefont {C.}~\bibnamefont {Toher}}, \bibinfo {author} {\bibfnamefont {R.}~\bibnamefont {Friedrich}}, \bibinfo {author} {\bibfnamefont {M.}~\bibnamefont {Esters}}, \bibinfo {author} {\bibfnamefont {M.~J.}\ \bibnamefont {Mehl}}, \bibinfo {author} {\bibfnamefont {A.~C.}\ \bibnamefont {Zettel}}, \bibinfo {author} {\bibfnamefont {Y.}~\bibnamefont {Lederer}}, \bibinfo {author} {\bibfnamefont {E.}~\bibnamefont {Zurek}}, \bibinfo {author} {\bibfnamefont {J.-P.}\ \bibnamefont {Maria}}, \bibinfo {author} {\bibfnamefont {D.~W.}\ \bibnamefont {Brenner}}, \bibinfo {author} {\bibfnamefont {X.}~\bibnamefont {Campilongo}}, \bibinfo {author} {\bibfnamefont {S.}~\bibnamefont {Filipovi{\'c}}}, \bibinfo {author} {\bibfnamefont {W.~G.}\
  \bibnamefont {Fahrenholtz}}, \bibinfo {author} {\bibfnamefont {C.~J.}\ \bibnamefont {Ryan}}, \bibinfo {author} {\bibfnamefont {C.~M.}\ \bibnamefont {DeSalle}}, \bibinfo {author} {\bibfnamefont {R.~J.}\ \bibnamefont {Crealese}}, \bibinfo {author} {\bibfnamefont {D.~E.}\ \bibnamefont {Wolfe}}, \bibinfo {author} {\bibfnamefont {A.}~\bibnamefont {Calzolari}},\ and\ \bibinfo {author} {\bibfnamefont {S.}~\bibnamefont {Curtarolo}},\ }\bibfield  {title} {\bibinfo {title} {Disordered enthalpy--entropy descriptor for high-entropy ceramics discovery},\ }\href@noop {} {\bibfield  {journal} {\bibinfo  {journal} {Nature}\ }\textbf {\bibinfo {volume} {625}},\ \bibinfo {pages} {66} (\bibinfo {year} {2024})}\BibitemShut {NoStop}%
\bibitem [{\citenamefont {Hedman}\ \emph {et~al.}(2022)\citenamefont {Hedman}, \citenamefont {Feltrin}, \citenamefont {Miyamoto},\ and\ \citenamefont {Akhtar}}]{Hedman2022}%
  \BibitemOpen
  \bibfield  {author} {\bibinfo {author} {\bibfnamefont {D.}~\bibnamefont {Hedman}}, \bibinfo {author} {\bibfnamefont {A.~C.}\ \bibnamefont {Feltrin}}, \bibinfo {author} {\bibfnamefont {Y.}~\bibnamefont {Miyamoto}},\ and\ \bibinfo {author} {\bibfnamefont {F.}~\bibnamefont {Akhtar}},\ }\bibfield  {title} {\bibinfo {title} {Ab initio aided design of novel quaternary, quinary and senary high-entropy borocarbides},\ }\href@noop {} {\bibfield  {journal} {\bibinfo  {journal} {Journal of Materials Science}\ }\textbf {\bibinfo {volume} {57}},\ \bibinfo {pages} {422} (\bibinfo {year} {2022})}\BibitemShut {NoStop}%
\bibitem [{\citenamefont {Qureshi}\ \emph {et~al.}(2025)\citenamefont {Qureshi}, \citenamefont {Wei}, \citenamefont {Liu}, \citenamefont {Paul}, \citenamefont {Kim}, \citenamefont {Zhang}, \citenamefont {Wang}, \citenamefont {Perepezko}, \citenamefont {Morgan},\ and\ \citenamefont {Szlufarska}}]{qureshi_predictive_2025}%
  \BibitemOpen
  \bibfield  {author} {\bibinfo {author} {\bibfnamefont {M.~W.}\ \bibnamefont {Qureshi}}, \bibinfo {author} {\bibfnamefont {S.}~\bibnamefont {Wei}}, \bibinfo {author} {\bibfnamefont {L.}~\bibnamefont {Liu}}, \bibinfo {author} {\bibfnamefont {S.}~\bibnamefont {Paul}}, \bibinfo {author} {\bibfnamefont {J.~Y.}\ \bibnamefont {Kim}}, \bibinfo {author} {\bibfnamefont {C.}~\bibnamefont {Zhang}}, \bibinfo {author} {\bibfnamefont {X.}~\bibnamefont {Wang}}, \bibinfo {author} {\bibfnamefont {J.~H.}\ \bibnamefont {Perepezko}}, \bibinfo {author} {\bibfnamefont {D.}~\bibnamefont {Morgan}},\ and\ \bibinfo {author} {\bibfnamefont {I.}~\bibnamefont {Szlufarska}},\ }\bibfield  {title} {\bibinfo {title} {Predictive screening of phase stability in high-entropy ceramics},\ }\href@noop {} {\bibfield  {journal} {\bibinfo  {journal} {Materials Advances}\ } (\bibinfo {year} {2025})}\BibitemShut {NoStop}%
\bibitem [{\citenamefont {Batatia}\ \emph {et~al.}(2022{\natexlab{a}})\citenamefont {Batatia}, \citenamefont {Batzner}, \citenamefont {Kov{\'a}cs}, \citenamefont {Musaelian}, \citenamefont {Simm}, \citenamefont {Drautz}, \citenamefont {Ortner}, \citenamefont {Kozinsky},\ and\ \citenamefont {Cs{\'a}nyi}}]{Batatia2022Design}%
  \BibitemOpen
  \bibfield  {author} {\bibinfo {author} {\bibfnamefont {I.}~\bibnamefont {Batatia}}, \bibinfo {author} {\bibfnamefont {S.}~\bibnamefont {Batzner}}, \bibinfo {author} {\bibfnamefont {D.~P.}\ \bibnamefont {Kov{\'a}cs}}, \bibinfo {author} {\bibfnamefont {A.}~\bibnamefont {Musaelian}}, \bibinfo {author} {\bibfnamefont {G.~N.~C.}\ \bibnamefont {Simm}}, \bibinfo {author} {\bibfnamefont {R.}~\bibnamefont {Drautz}}, \bibinfo {author} {\bibfnamefont {C.}~\bibnamefont {Ortner}}, \bibinfo {author} {\bibfnamefont {B.}~\bibnamefont {Kozinsky}},\ and\ \bibinfo {author} {\bibfnamefont {G.}~\bibnamefont {Cs{\'a}nyi}},\ }\href {https://doi.org/10.48550/arXiv.2205.06643} {\bibinfo {title} {The design space of e(3)-equivariant atom-centered interatomic potentials}} (\bibinfo {year} {2022}{\natexlab{a}}),\ \Eprint {https://arxiv.org/abs/2205.06643} {arXiv:2205.06643} \BibitemShut {NoStop}%
\bibitem [{\citenamefont {Batatia}\ \emph {et~al.}(2022{\natexlab{b}})\citenamefont {Batatia}, \citenamefont {Kovacs}, \citenamefont {Simm}, \citenamefont {Ortner},\ and\ \citenamefont {Csanyi}}]{Batatia2022mace}%
  \BibitemOpen
  \bibfield  {author} {\bibinfo {author} {\bibfnamefont {I.}~\bibnamefont {Batatia}}, \bibinfo {author} {\bibfnamefont {D.~P.}\ \bibnamefont {Kovacs}}, \bibinfo {author} {\bibfnamefont {G.~N.~C.}\ \bibnamefont {Simm}}, \bibinfo {author} {\bibfnamefont {C.}~\bibnamefont {Ortner}},\ and\ \bibinfo {author} {\bibfnamefont {G.}~\bibnamefont {Csanyi}},\ }\bibfield  {title} {\bibinfo {title} {{MACE}: Higher order equivariant message passing neural networks for fast and accurate force fields},\ }in\ \href@noop {} {\emph {\bibinfo {booktitle} {Advances in Neural Information Processing Systems}}},\ \bibinfo {editor} {edited by\ \bibinfo {editor} {\bibfnamefont {A.~H.}\ \bibnamefont {Oh}}, \bibinfo {editor} {\bibfnamefont {A.}~\bibnamefont {Agarwal}}, \bibinfo {editor} {\bibfnamefont {D.}~\bibnamefont {Belgrave}},\ and\ \bibinfo {editor} {\bibfnamefont {K.}~\bibnamefont {Cho}}}\ (\bibinfo {year} {2022})\BibitemShut {NoStop}%
\bibitem [{\citenamefont {Drautz}(2019)}]{drautz_atomic_2019}%
  \BibitemOpen
  \bibfield  {author} {\bibinfo {author} {\bibfnamefont {R.}~\bibnamefont {Drautz}},\ }\bibfield  {title} {\bibinfo {title} {Atomic cluster expansion for accurate and transferable interatomic potentials},\ }\href@noop {} {\bibfield  {journal} {\bibinfo  {journal} {Physical Review B}\ }\textbf {\bibinfo {volume} {99}},\ \bibinfo {pages} {014104} (\bibinfo {year} {2019})}\BibitemShut {NoStop}%
\bibitem [{\citenamefont {Kov{\'a}cs}\ \emph {et~al.}(2023)\citenamefont {Kov{\'a}cs}, \citenamefont {Batatia}, \citenamefont {Arany},\ and\ \citenamefont {Cs{\'a}nyi}}]{Kovacs2023}%
  \BibitemOpen
  \bibfield  {author} {\bibinfo {author} {\bibfnamefont {D.~P.}\ \bibnamefont {Kov{\'a}cs}}, \bibinfo {author} {\bibfnamefont {I.}~\bibnamefont {Batatia}}, \bibinfo {author} {\bibfnamefont {E.~S.}\ \bibnamefont {Arany}},\ and\ \bibinfo {author} {\bibfnamefont {G.}~\bibnamefont {Cs{\'a}nyi}},\ }\bibfield  {title} {\bibinfo {title} {Evaluation of the mace force field architecture: From medicinal chemistry to materials science},\ }\href@noop {} {\bibfield  {journal} {\bibinfo  {journal} {The Journal of Chemical Physics}\ }\textbf {\bibinfo {volume} {159}},\ \bibinfo {pages} {044118} (\bibinfo {year} {2023})}\BibitemShut {NoStop}%
\bibitem [{\citenamefont {Kresse}\ and\ \citenamefont {Furthm{\"u}ller}(1996)}]{Kresse199611169}%
  \BibitemOpen
  \bibfield  {author} {\bibinfo {author} {\bibfnamefont {G.}~\bibnamefont {Kresse}}\ and\ \bibinfo {author} {\bibfnamefont {J.}~\bibnamefont {Furthm{\"u}ller}},\ }\bibfield  {title} {\bibinfo {title} {Efficient iterative schemes for ab initio total-energy calculations using a plane-wave basis set},\ }\href@noop {} {\bibfield  {journal} {\bibinfo  {journal} {Physical Review B - Condensed Matter and Materials Physics}\ }\textbf {\bibinfo {volume} {54}},\ \bibinfo {pages} {11169 } (\bibinfo {year} {1996})}\BibitemShut {NoStop}%
\bibitem [{\citenamefont {Perdew}\ \emph {et~al.}(1996)\citenamefont {Perdew}, \citenamefont {Burke},\ and\ \citenamefont {Ernzerhof}}]{Perdew1996}%
  \BibitemOpen
  \bibfield  {author} {\bibinfo {author} {\bibfnamefont {J.~P.}\ \bibnamefont {Perdew}}, \bibinfo {author} {\bibfnamefont {K.}~\bibnamefont {Burke}},\ and\ \bibinfo {author} {\bibfnamefont {M.}~\bibnamefont {Ernzerhof}},\ }\bibfield  {title} {\bibinfo {title} {Generalized gradient approximation made simple},\ }\href@noop {} {\bibfield  {journal} {\bibinfo  {journal} {Phys. Rev. Lett.}\ }\textbf {\bibinfo {volume} {77}},\ \bibinfo {pages} {3865} (\bibinfo {year} {1996})}\BibitemShut {NoStop}%
\bibitem [{\citenamefont {Kresse}(1999)}]{Kresse19991758}%
  \BibitemOpen
  \bibfield  {author} {\bibinfo {author} {\bibfnamefont {G.}~\bibnamefont {Kresse}},\ }\bibfield  {title} {\bibinfo {title} {From ultrasoft pseudopotentials to the projector augmented-wave method},\ }\href@noop {} {\bibfield  {journal} {\bibinfo  {journal} {Physical Review B - Condensed Matter and Materials Physics}\ }\textbf {\bibinfo {volume} {59}},\ \bibinfo {pages} {1758 } (\bibinfo {year} {1999})}\BibitemShut {NoStop}%
\bibitem [{\citenamefont {Methfessel}\ and\ \citenamefont {Paxton}(1989)}]{Methfessel19893616}%
  \BibitemOpen
  \bibfield  {author} {\bibinfo {author} {\bibfnamefont {M.}~\bibnamefont {Methfessel}}\ and\ \bibinfo {author} {\bibfnamefont {A.}~\bibnamefont {Paxton}},\ }\bibfield  {title} {\bibinfo {title} {High-precision sampling for brillouin-zone integration in metals},\ }\href@noop {} {\bibfield  {journal} {\bibinfo  {journal} {Physical Review B}\ }\textbf {\bibinfo {volume} {40}},\ \bibinfo {pages} {3616 } (\bibinfo {year} {1989})}\BibitemShut {NoStop}%
\bibitem [{\citenamefont {von Helmholtz}(1882)}]{Helmholtz1882}%
  \BibitemOpen
  \bibfield  {author} {\bibinfo {author} {\bibfnamefont {H.}~\bibnamefont {von Helmholtz}},\ }\href@noop {} {\emph {\bibinfo {title} {Die Thermodynamik chemischer Vorg{\"a}nge}}}\ (\bibinfo  {publisher} {Berichte der K{\"o}niglichen Preussischen Akademie der Wissenschaften zu Berlin},\ \bibinfo {address} {Berlin},\ \bibinfo {year} {1882})\ p.~\bibinfo {pages} {22}\BibitemShut {NoStop}%
\bibitem [{\citenamefont {M{\"u}ller}\ and\ \citenamefont {Natarajan}(2024)}]{muller_first-principles_2024}%
  \BibitemOpen
  \bibfield  {author} {\bibinfo {author} {\bibfnamefont {Y.~L.}\ \bibnamefont {M{\"u}ller}}\ and\ \bibinfo {author} {\bibfnamefont {A.~R.}\ \bibnamefont {Natarajan}},\ }\bibfield  {title} {\bibinfo {title} {First-principles thermodynamics of precipitation in aluminum-containing refractory alloys},\ }\href@noop {} {\bibfield  {journal} {\bibinfo  {journal} {Acta Materialia}\ }\textbf {\bibinfo {volume} {274}},\ \bibinfo {pages} {119995} (\bibinfo {year} {2024})}\BibitemShut {NoStop}%
\bibitem [{\citenamefont {Wei}\ \emph {et~al.}(2026)\citenamefont {Wei}, \citenamefont {Qureshi}, \citenamefont {Wei}, \citenamefont {Liu}, \citenamefont {Hu}, \citenamefont {Xi}, \citenamefont {Attarian}, \citenamefont {Su}, \citenamefont {Zhang}, \citenamefont {Willing}, \citenamefont {Wang}, \citenamefont {Sridharan}, \citenamefont {Voyles}, \citenamefont {Perepezko},\ and\ \citenamefont {Szlufarska}}]{Wei2026aa}%
  \BibitemOpen
  \bibfield  {author} {\bibinfo {author} {\bibfnamefont {S.}~\bibnamefont {Wei}}, \bibinfo {author} {\bibfnamefont {M.~W.}\ \bibnamefont {Qureshi}}, \bibinfo {author} {\bibfnamefont {J.}~\bibnamefont {Wei}}, \bibinfo {author} {\bibfnamefont {L.}~\bibnamefont {Liu}}, \bibinfo {author} {\bibfnamefont {X.}~\bibnamefont {Hu}}, \bibinfo {author} {\bibfnamefont {J.}~\bibnamefont {Xi}}, \bibinfo {author} {\bibfnamefont {S.}~\bibnamefont {Attarian}}, \bibinfo {author} {\bibfnamefont {R.}~\bibnamefont {Su}}, \bibinfo {author} {\bibfnamefont {H.}~\bibnamefont {Zhang}}, \bibinfo {author} {\bibfnamefont {E.}~\bibnamefont {Willing}}, \bibinfo {author} {\bibfnamefont {X.}~\bibnamefont {Wang}}, \bibinfo {author} {\bibfnamefont {K.}~\bibnamefont {Sridharan}}, \bibinfo {author} {\bibfnamefont {P.~M.}\ \bibnamefont {Voyles}}, \bibinfo {author} {\bibfnamefont {J.~H.}\ \bibnamefont {Perepezko}},\ and\ \bibinfo {author} {\bibfnamefont {I.}~\bibnamefont {Szlufarska}},\ }\bibfield  {title} {\bibinfo {title} {Short-range order in
  high entropy carbides},\ }\href@noop {} {\bibfield  {journal} {\bibinfo  {journal} {Nature Communications}\ }\textbf {\bibinfo {volume} {17}},\ \bibinfo {pages} {2362} (\bibinfo {year} {2026})}\BibitemShut {NoStop}%
\bibitem [{\citenamefont {Zunger}\ \emph {et~al.}(1990)\citenamefont {Zunger}, \citenamefont {Wei}, \citenamefont {Ferreira},\ and\ \citenamefont {Bernard}}]{zunger1990special}%
  \BibitemOpen
  \bibfield  {author} {\bibinfo {author} {\bibfnamefont {A.}~\bibnamefont {Zunger}}, \bibinfo {author} {\bibfnamefont {S.-H.}\ \bibnamefont {Wei}}, \bibinfo {author} {\bibfnamefont {L.~G.}\ \bibnamefont {Ferreira}},\ and\ \bibinfo {author} {\bibfnamefont {J.~E.}\ \bibnamefont {Bernard}},\ }\bibfield  {title} {\bibinfo {title} {Special quasirandom structures},\ }\href@noop {} {\bibfield  {journal} {\bibinfo  {journal} {Physical review letters}\ }\textbf {\bibinfo {volume} {65}},\ \bibinfo {pages} {353} (\bibinfo {year} {1990})}\BibitemShut {NoStop}%
\bibitem [{\citenamefont {Thomas}\ and\ \citenamefont {Ven}(2013)}]{thomas2013finite}%
  \BibitemOpen
  \bibfield  {author} {\bibinfo {author} {\bibfnamefont {J.~C.}\ \bibnamefont {Thomas}}\ and\ \bibinfo {author} {\bibfnamefont {A.~V.~d.}\ \bibnamefont {Ven}},\ }\bibfield  {title} {\bibinfo {title} {Finite-temperature properties of strongly anharmonic and mechanically unstable crystal phases from first principles},\ }\href@noop {} {\bibfield  {journal} {\bibinfo  {journal} {Physical Review B}\ }\textbf {\bibinfo {volume} {88}},\ \bibinfo {pages} {214111} (\bibinfo {year} {2013})}\BibitemShut {NoStop}%
\bibitem [{\citenamefont {Puchala}\ and\ \citenamefont {Van~der Ven}(2013)}]{puchala2013thermodynamics}%
  \BibitemOpen
  \bibfield  {author} {\bibinfo {author} {\bibfnamefont {B.}~\bibnamefont {Puchala}}\ and\ \bibinfo {author} {\bibfnamefont {A.}~\bibnamefont {Van~der Ven}},\ }\bibfield  {title} {\bibinfo {title} {Thermodynamics of the zr-o system from first-principles calculations},\ }\href@noop {} {\bibfield  {journal} {\bibinfo  {journal} {Physical Review B---Condensed Matter and Materials Physics}\ }\textbf {\bibinfo {volume} {88}},\ \bibinfo {pages} {094108} (\bibinfo {year} {2013})}\BibitemShut {NoStop}%
\bibitem [{\citenamefont {Van~der Ven}\ \emph {et~al.}(2018)\citenamefont {Van~der Ven}, \citenamefont {Thomas}, \citenamefont {Puchala},\ and\ \citenamefont {Natarajan}}]{van2018first}%
  \BibitemOpen
  \bibfield  {author} {\bibinfo {author} {\bibfnamefont {A.}~\bibnamefont {Van~der Ven}}, \bibinfo {author} {\bibfnamefont {J.~C.}\ \bibnamefont {Thomas}}, \bibinfo {author} {\bibfnamefont {B.}~\bibnamefont {Puchala}},\ and\ \bibinfo {author} {\bibfnamefont {A.~R.}\ \bibnamefont {Natarajan}},\ }\bibfield  {title} {\bibinfo {title} {First-principles statistical mechanics of multicomponent crystals},\ }\href@noop {} {\bibfield  {journal} {\bibinfo  {journal} {Annual Review of Materials Research}\ }\textbf {\bibinfo {volume} {48}},\ \bibinfo {pages} {27} (\bibinfo {year} {2018})}\BibitemShut {NoStop}%
\bibitem [{\citenamefont {Watkins}\ \emph {et~al.}(2024)\citenamefont {Watkins}, \citenamefont {Haas~Blacksher}, \citenamefont {Stubbers}, \citenamefont {Thompson},\ and\ \citenamefont {Weinberger}}]{watkins_insights_2024}%
  \BibitemOpen
  \bibfield  {author} {\bibinfo {author} {\bibfnamefont {B.~R.}\ \bibnamefont {Watkins}}, \bibinfo {author} {\bibfnamefont {C.}~\bibnamefont {Haas~Blacksher}}, \bibinfo {author} {\bibfnamefont {A.}~\bibnamefont {Stubbers}}, \bibinfo {author} {\bibfnamefont {G.~B.}\ \bibnamefont {Thompson}},\ and\ \bibinfo {author} {\bibfnamefont {C.~R.}\ \bibnamefont {Weinberger}},\ }\bibfield  {title} {\bibinfo {title} {Insights into the anomalous hardness of the tantalum carbides from dislocation mobility},\ }\href@noop {} {\bibfield  {journal} {\bibinfo  {journal} {Nature Communications}\ }\textbf {\bibinfo {volume} {15}},\ \bibinfo {pages} {10585} (\bibinfo {year} {2024})}\BibitemShut {NoStop}%
\bibitem [{\citenamefont {Watkins}\ \emph {et~al.}(2025)\citenamefont {Watkins}, \citenamefont {Huang}, \citenamefont {Stubbers}, \citenamefont {Thompson},\ and\ \citenamefont {Weinberger}}]{WATKINS2025121350}%
  \BibitemOpen
  \bibfield  {author} {\bibinfo {author} {\bibfnamefont {B.~R.}\ \bibnamefont {Watkins}}, \bibinfo {author} {\bibfnamefont {Y.}~\bibnamefont {Huang}}, \bibinfo {author} {\bibfnamefont {A.}~\bibnamefont {Stubbers}}, \bibinfo {author} {\bibfnamefont {G.~B.}\ \bibnamefont {Thompson}},\ and\ \bibinfo {author} {\bibfnamefont {C.~R.}\ \bibnamefont {Weinberger}},\ }\bibfield  {title} {\bibinfo {title} {Plasticity-fracture competition and anomalous hardness in the hard metals},\ }\href@noop {} {\bibfield  {journal} {\bibinfo  {journal} {Acta Materialia}\ }\textbf {\bibinfo {volume} {298}},\ \bibinfo {pages} {121350} (\bibinfo {year} {2025})}\BibitemShut {NoStop}%
\bibitem [{\citenamefont {Yu}\ \emph {et~al.}(2017)\citenamefont {Yu}, \citenamefont {Bahadori}, \citenamefont {Thompson},\ and\ \citenamefont {Weinberger}}]{Hang2017}%
  \BibitemOpen
  \bibfield  {author} {\bibinfo {author} {\bibfnamefont {H.}~\bibnamefont {Yu}}, \bibinfo {author} {\bibfnamefont {M.}~\bibnamefont {Bahadori}}, \bibinfo {author} {\bibfnamefont {G.~B.}\ \bibnamefont {Thompson}},\ and\ \bibinfo {author} {\bibfnamefont {C.~R.}\ \bibnamefont {Weinberger}},\ }\bibfield  {title} {\bibinfo {title} {Understanding dislocation slip in stoichiometric rocksalt transition metal carbides and nitrides},\ }\href@noop {} {\bibfield  {journal} {\bibinfo  {journal} {Journal of Materials Science}\ }\textbf {\bibinfo {volume} {52}},\ \bibinfo {pages} {6235} (\bibinfo {year} {2017})}\BibitemShut {NoStop}%
\bibitem [{\citenamefont {Zhang}\ \emph {et~al.}(2024)\citenamefont {Zhang}, \citenamefont {He}, \citenamefont {Xiong}, \citenamefont {Huang}, \citenamefont {Xu}, \citenamefont {Ma}, \citenamefont {Xiang}, \citenamefont {Fu}, \citenamefont {Kai}, \citenamefont {Wu},\ and\ \citenamefont {Zhao}}]{Zhang:2024aa}%
  \BibitemOpen
  \bibfield  {author} {\bibinfo {author} {\bibfnamefont {J.}~\bibnamefont {Zhang}}, \bibinfo {author} {\bibfnamefont {L.}~\bibnamefont {He}}, \bibinfo {author} {\bibfnamefont {Y.}~\bibnamefont {Xiong}}, \bibinfo {author} {\bibfnamefont {S.}~\bibnamefont {Huang}}, \bibinfo {author} {\bibfnamefont {B.}~\bibnamefont {Xu}}, \bibinfo {author} {\bibfnamefont {S.}~\bibnamefont {Ma}}, \bibinfo {author} {\bibfnamefont {X.}~\bibnamefont {Xiang}}, \bibinfo {author} {\bibfnamefont {H.}~\bibnamefont {Fu}}, \bibinfo {author} {\bibfnamefont {J.}~\bibnamefont {Kai}}, \bibinfo {author} {\bibfnamefont {Z.}~\bibnamefont {Wu}},\ and\ \bibinfo {author} {\bibfnamefont {S.}~\bibnamefont {Zhao}},\ }\bibfield  {title} {\bibinfo {title} {Local-distortion-informed exceptional multicomponent transition-metal carbides uncovered by machine learning},\ }\href@noop {} {\bibfield  {journal} {\bibinfo  {journal} {npj Computational Materials}\ }\textbf {\bibinfo {volume} {10}},\ \bibinfo {pages} {162} (\bibinfo {year} {2024})}\BibitemShut
  {NoStop}%
\bibitem [{\citenamefont {Mazhnik}\ and\ \citenamefont {Oganov}(2019)}]{mazhnik_model_2019}%
  \BibitemOpen
  \bibfield  {author} {\bibinfo {author} {\bibfnamefont {E.}~\bibnamefont {Mazhnik}}\ and\ \bibinfo {author} {\bibfnamefont {A.~R.}\ \bibnamefont {Oganov}},\ }\bibfield  {title} {\bibinfo {title} {A model of hardness and fracture toughness of solids},\ }\href@noop {} {\bibfield  {journal} {\bibinfo  {journal} {Journal of Applied Physics}\ }\textbf {\bibinfo {volume} {126}},\ \bibinfo {pages} {125109} (\bibinfo {year} {2019})}\BibitemShut {NoStop}%
\bibitem [{\citenamefont {Teter}(1998)}]{Teter1998}%
  \BibitemOpen
  \bibfield  {author} {\bibinfo {author} {\bibfnamefont {D.~M.}\ \bibnamefont {Teter}},\ }\bibfield  {title} {\bibinfo {title} {Computational alchemy: The search for new superhard materials},\ }\href@noop {} {\bibfield  {journal} {\bibinfo  {journal} {MRS Bulletin}\ }\textbf {\bibinfo {volume} {23}},\ \bibinfo {pages} {22} (\bibinfo {year} {1998})}\BibitemShut {NoStop}%
\bibitem [{\citenamefont {Chen}\ \emph {et~al.}(2011)\citenamefont {Chen}, \citenamefont {Niu}, \citenamefont {Li},\ and\ \citenamefont {Li}}]{CHEN20111275}%
  \BibitemOpen
  \bibfield  {author} {\bibinfo {author} {\bibfnamefont {X.-Q.}\ \bibnamefont {Chen}}, \bibinfo {author} {\bibfnamefont {H.}~\bibnamefont {Niu}}, \bibinfo {author} {\bibfnamefont {D.}~\bibnamefont {Li}},\ and\ \bibinfo {author} {\bibfnamefont {Y.}~\bibnamefont {Li}},\ }\bibfield  {title} {\bibinfo {title} {Modeling hardness of polycrystalline materials and bulk metallic glasses},\ }\href@noop {} {\bibfield  {journal} {\bibinfo  {journal} {Intermetallics}\ }\textbf {\bibinfo {volume} {19}},\ \bibinfo {pages} {1275} (\bibinfo {year} {2011})}\BibitemShut {NoStop}%
\bibitem [{\citenamefont {Tian}\ \emph {et~al.}(2012)\citenamefont {Tian}, \citenamefont {Xu},\ and\ \citenamefont {Zhao}}]{TIAN201293}%
  \BibitemOpen
  \bibfield  {author} {\bibinfo {author} {\bibfnamefont {Y.}~\bibnamefont {Tian}}, \bibinfo {author} {\bibfnamefont {B.}~\bibnamefont {Xu}},\ and\ \bibinfo {author} {\bibfnamefont {Z.}~\bibnamefont {Zhao}},\ }\bibfield  {title} {\bibinfo {title} {Microscopic theory of hardness and design of novel superhard crystals},\ }\href@noop {} {\bibfield  {journal} {\bibinfo  {journal} {International Journal of Refractory Metals and Hard Materials}\ }\textbf {\bibinfo {volume} {33}},\ \bibinfo {pages} {93} (\bibinfo {year} {2012})}\BibitemShut {NoStop}%
\bibitem [{\citenamefont {Deng}\ \emph {et~al.}(2024)\citenamefont {Deng}, \citenamefont {Choi}, \citenamefont {Zhong}, \citenamefont {Riebesell}, \citenamefont {Anand}, \citenamefont {Li}, \citenamefont {Jun}, \citenamefont {Persson},\ and\ \citenamefont {Ceder}}]{deng2024}%
  \BibitemOpen
  \bibfield  {author} {\bibinfo {author} {\bibfnamefont {B.}~\bibnamefont {Deng}}, \bibinfo {author} {\bibfnamefont {Y.}~\bibnamefont {Choi}}, \bibinfo {author} {\bibfnamefont {P.}~\bibnamefont {Zhong}}, \bibinfo {author} {\bibfnamefont {J.}~\bibnamefont {Riebesell}}, \bibinfo {author} {\bibfnamefont {S.}~\bibnamefont {Anand}}, \bibinfo {author} {\bibfnamefont {Z.}~\bibnamefont {Li}}, \bibinfo {author} {\bibfnamefont {K.}~\bibnamefont {Jun}}, \bibinfo {author} {\bibfnamefont {K.~A.}\ \bibnamefont {Persson}},\ and\ \bibinfo {author} {\bibfnamefont {G.}~\bibnamefont {Ceder}},\ }\href {https://arxiv.org/abs/2405.07105} {\bibinfo {title} {Overcoming systematic softening in universal machine learning interatomic potentials by fine-tuning}} (\bibinfo {year} {2024}),\ \Eprint {https://arxiv.org/abs/2405.07105} {arXiv:2405.07105 [cond-mat.mtrl-sci]} \BibitemShut {NoStop}%
\bibitem [{\citenamefont {Kaur}\ \emph {et~al.}(2025)\citenamefont {Kaur}, \citenamefont {Della~Pia}, \citenamefont {Batatia}, \citenamefont {Advincula}, \citenamefont {Shi}, \citenamefont {Lan}, \citenamefont {Cs{\'a}nyi}, \citenamefont {Michaelides},\ and\ \citenamefont {Kapil}}]{D4FD00107A}%
  \BibitemOpen
  \bibfield  {author} {\bibinfo {author} {\bibfnamefont {H.}~\bibnamefont {Kaur}}, \bibinfo {author} {\bibfnamefont {F.}~\bibnamefont {Della~Pia}}, \bibinfo {author} {\bibfnamefont {I.}~\bibnamefont {Batatia}}, \bibinfo {author} {\bibfnamefont {X.~R.}\ \bibnamefont {Advincula}}, \bibinfo {author} {\bibfnamefont {B.~X.}\ \bibnamefont {Shi}}, \bibinfo {author} {\bibfnamefont {J.}~\bibnamefont {Lan}}, \bibinfo {author} {\bibfnamefont {G.}~\bibnamefont {Cs{\'a}nyi}}, \bibinfo {author} {\bibfnamefont {A.}~\bibnamefont {Michaelides}},\ and\ \bibinfo {author} {\bibfnamefont {V.}~\bibnamefont {Kapil}},\ }\bibfield  {title} {\bibinfo {title} {Data-efficient fine-tuning of foundational models for first-principles quality sublimation enthalpies},\ }\href@noop {} {\bibfield  {journal} {\bibinfo  {journal} {Faraday Discuss.}\ }\textbf {\bibinfo {volume} {256}},\ \bibinfo {pages} {120} (\bibinfo {year} {2025})}\BibitemShut {NoStop}%
\bibitem [{\citenamefont {Radova}\ \emph {et~al.}(2025)\citenamefont {Radova}, \citenamefont {Stark}, \citenamefont {Allen}, \citenamefont {Maurer},\ and\ \citenamefont {Bart{\'o}k}}]{Radova2025}%
  \BibitemOpen
  \bibfield  {author} {\bibinfo {author} {\bibfnamefont {M.}~\bibnamefont {Radova}}, \bibinfo {author} {\bibfnamefont {W.~G.}\ \bibnamefont {Stark}}, \bibinfo {author} {\bibfnamefont {C.~S.}\ \bibnamefont {Allen}}, \bibinfo {author} {\bibfnamefont {R.~J.}\ \bibnamefont {Maurer}},\ and\ \bibinfo {author} {\bibfnamefont {A.~P.}\ \bibnamefont {Bart{\'o}k}},\ }\bibfield  {title} {\bibinfo {title} {Fine-tuning foundation models of materials interatomic potentials with frozen transfer learning},\ }\href@noop {} {\bibfield  {journal} {\bibinfo  {journal} {npj Computational Materials}\ }\textbf {\bibinfo {volume} {11}},\ \bibinfo {pages} {237} (\bibinfo {year} {2025})}\BibitemShut {NoStop}%
\bibitem [{\citenamefont {Piersante}\ and\ \citenamefont {Natarajan}(2026)}]{piersante}%
  \BibitemOpen
  \bibfield  {author} {\bibinfo {author} {\bibfnamefont {L.}~\bibnamefont {Piersante}}\ and\ \bibinfo {author} {\bibfnamefont {A.~R.}\ \bibnamefont {Natarajan}},\ }\href {https://arxiv.org/abs/2601.12984} {\bibinfo {title} {Machine learning interatomic potentials for solid-state precipitation}} (\bibinfo {year} {2026}),\ \Eprint {https://arxiv.org/abs/2601.12984} {arXiv:2601.12984 [cond-mat.mtrl-sci]} \BibitemShut {NoStop}%
\bibitem [{\citenamefont {Natarajan}\ \emph {et~al.}(2020)\citenamefont {Natarajan}, \citenamefont {Dolin},\ and\ \citenamefont {{Van der Ven}}}]{NATARAJAN2020171}%
  \BibitemOpen
  \bibfield  {author} {\bibinfo {author} {\bibfnamefont {A.~R.}\ \bibnamefont {Natarajan}}, \bibinfo {author} {\bibfnamefont {P.}~\bibnamefont {Dolin}},\ and\ \bibinfo {author} {\bibfnamefont {A.}~\bibnamefont {{Van der Ven}}},\ }\bibfield  {title} {\bibinfo {title} {Crystallography, thermodynamics and phase transitions in refractory binary alloys},\ }\href@noop {} {\bibfield  {journal} {\bibinfo  {journal} {Acta Materialia}\ }\textbf {\bibinfo {volume} {200}},\ \bibinfo {pages} {171} (\bibinfo {year} {2020})}\BibitemShut {NoStop}%
\bibitem [{\citenamefont {Hossain}\ \emph {et~al.}(2021)\citenamefont {Hossain}, \citenamefont {Borman}, \citenamefont {Oses}, \citenamefont {Esters}, \citenamefont {Toher}, \citenamefont {Feng}, \citenamefont {Kumar}, \citenamefont {Fahrenholtz}, \citenamefont {Curtarolo}, \citenamefont {Brenner} \emph {et~al.}}]{hossain2021entropy}%
  \BibitemOpen
  \bibfield  {author} {\bibinfo {author} {\bibfnamefont {M.~D.}\ \bibnamefont {Hossain}}, \bibinfo {author} {\bibfnamefont {T.}~\bibnamefont {Borman}}, \bibinfo {author} {\bibfnamefont {C.}~\bibnamefont {Oses}}, \bibinfo {author} {\bibfnamefont {M.}~\bibnamefont {Esters}}, \bibinfo {author} {\bibfnamefont {C.}~\bibnamefont {Toher}}, \bibinfo {author} {\bibfnamefont {L.}~\bibnamefont {Feng}}, \bibinfo {author} {\bibfnamefont {A.}~\bibnamefont {Kumar}}, \bibinfo {author} {\bibfnamefont {W.~G.}\ \bibnamefont {Fahrenholtz}}, \bibinfo {author} {\bibfnamefont {S.}~\bibnamefont {Curtarolo}}, \bibinfo {author} {\bibfnamefont {D.}~\bibnamefont {Brenner}}, \emph {et~al.},\ }\bibfield  {title} {\bibinfo {title} {Entropy landscaping of high-entropy carbides},\ }\href@noop {} {\bibfield  {journal} {\bibinfo  {journal} {Advanced Materials}\ }\textbf {\bibinfo {volume} {33}},\ \bibinfo {pages} {2102904} (\bibinfo {year} {2021})}\BibitemShut {NoStop}%
\bibitem [{\citenamefont {Castle}\ \emph {et~al.}(2018)\citenamefont {Castle}, \citenamefont {Csan{\'a}di}, \citenamefont {Grasso}, \citenamefont {Dusza},\ and\ \citenamefont {Reece}}]{castle2018processing}%
  \BibitemOpen
  \bibfield  {author} {\bibinfo {author} {\bibfnamefont {E.}~\bibnamefont {Castle}}, \bibinfo {author} {\bibfnamefont {T.}~\bibnamefont {Csan{\'a}di}}, \bibinfo {author} {\bibfnamefont {S.}~\bibnamefont {Grasso}}, \bibinfo {author} {\bibfnamefont {J.}~\bibnamefont {Dusza}},\ and\ \bibinfo {author} {\bibfnamefont {M.}~\bibnamefont {Reece}},\ }\bibfield  {title} {\bibinfo {title} {Processing and properties of high-entropy ultra-high temperature carbides},\ }\href@noop {} {\bibfield  {journal} {\bibinfo  {journal} {Scientific reports}\ }\textbf {\bibinfo {volume} {8}},\ \bibinfo {pages} {8609} (\bibinfo {year} {2018})}\BibitemShut {NoStop}%
\bibitem [{\citenamefont {Harrington}\ \emph {et~al.}(2019)\citenamefont {Harrington}, \citenamefont {Gild}, \citenamefont {Sarker}, \citenamefont {Toher}, \citenamefont {Rost}, \citenamefont {Dippo}, \citenamefont {McElfresh}, \citenamefont {Kaufmann}, \citenamefont {Marin}, \citenamefont {Borowski}, \citenamefont {Hopkins}, \citenamefont {Luo}, \citenamefont {Curtarolo}, \citenamefont {Brenner},\ and\ \citenamefont {Vecchio}}]{HARRINGTON2019271}%
  \BibitemOpen
  \bibfield  {author} {\bibinfo {author} {\bibfnamefont {T.~J.}\ \bibnamefont {Harrington}}, \bibinfo {author} {\bibfnamefont {J.}~\bibnamefont {Gild}}, \bibinfo {author} {\bibfnamefont {P.}~\bibnamefont {Sarker}}, \bibinfo {author} {\bibfnamefont {C.}~\bibnamefont {Toher}}, \bibinfo {author} {\bibfnamefont {C.~M.}\ \bibnamefont {Rost}}, \bibinfo {author} {\bibfnamefont {O.~F.}\ \bibnamefont {Dippo}}, \bibinfo {author} {\bibfnamefont {C.}~\bibnamefont {McElfresh}}, \bibinfo {author} {\bibfnamefont {K.}~\bibnamefont {Kaufmann}}, \bibinfo {author} {\bibfnamefont {E.}~\bibnamefont {Marin}}, \bibinfo {author} {\bibfnamefont {L.}~\bibnamefont {Borowski}}, \bibinfo {author} {\bibfnamefont {P.~E.}\ \bibnamefont {Hopkins}}, \bibinfo {author} {\bibfnamefont {J.}~\bibnamefont {Luo}}, \bibinfo {author} {\bibfnamefont {S.}~\bibnamefont {Curtarolo}}, \bibinfo {author} {\bibfnamefont {D.~W.}\ \bibnamefont {Brenner}},\ and\ \bibinfo {author} {\bibfnamefont {K.~S.}\ \bibnamefont {Vecchio}},\ }\bibfield  {title} {\bibinfo
  {title} {Phase stability and mechanical properties of novel high entropy transition metal carbides},\ }\href@noop {} {\bibfield  {journal} {\bibinfo  {journal} {Acta Materialia}\ }\textbf {\bibinfo {volume} {166}},\ \bibinfo {pages} {271} (\bibinfo {year} {2019})}\BibitemShut {NoStop}%
\bibitem [{\citenamefont {Chicardi}\ \emph {et~al.}(2019)\citenamefont {Chicardi}, \citenamefont {Garc{\'\i}a-Garrido},\ and\ \citenamefont {Gotor}}]{chicardi2019low}%
  \BibitemOpen
  \bibfield  {author} {\bibinfo {author} {\bibfnamefont {E.}~\bibnamefont {Chicardi}}, \bibinfo {author} {\bibfnamefont {C.}~\bibnamefont {Garc{\'\i}a-Garrido}},\ and\ \bibinfo {author} {\bibfnamefont {F.}~\bibnamefont {Gotor}},\ }\bibfield  {title} {\bibinfo {title} {Low temperature synthesis of an equiatomic (tizrhfvnb) c5 high entropy carbide by a mechanically-induced carbon diffusion route},\ }\href@noop {} {\bibfield  {journal} {\bibinfo  {journal} {Ceramics International}\ }\textbf {\bibinfo {volume} {45}},\ \bibinfo {pages} {21858} (\bibinfo {year} {2019})}\BibitemShut {NoStop}%
\bibitem [{\citenamefont {Wei}\ \emph {et~al.}(2019)\citenamefont {Wei}, \citenamefont {Liu}, \citenamefont {Li}, \citenamefont {Qin}, \citenamefont {Liang},\ and\ \citenamefont {Zhang}}]{wei2019high}%
  \BibitemOpen
  \bibfield  {author} {\bibinfo {author} {\bibfnamefont {X.-F.}\ \bibnamefont {Wei}}, \bibinfo {author} {\bibfnamefont {J.-X.}\ \bibnamefont {Liu}}, \bibinfo {author} {\bibfnamefont {F.}~\bibnamefont {Li}}, \bibinfo {author} {\bibfnamefont {Y.}~\bibnamefont {Qin}}, \bibinfo {author} {\bibfnamefont {Y.-C.}\ \bibnamefont {Liang}},\ and\ \bibinfo {author} {\bibfnamefont {G.-J.}\ \bibnamefont {Zhang}},\ }\bibfield  {title} {\bibinfo {title} {High entropy carbide ceramics from different starting materials},\ }\href@noop {} {\bibfield  {journal} {\bibinfo  {journal} {Journal of the European Ceramic Society}\ }\textbf {\bibinfo {volume} {39}},\ \bibinfo {pages} {2989} (\bibinfo {year} {2019})}\BibitemShut {NoStop}%
\bibitem [{\citenamefont {Chicardi}\ \emph {et~al.}(2020)\citenamefont {Chicardi}, \citenamefont {Garc{\'\i}a-Garrido}, \citenamefont {Hern{\'a}ndez-Saz},\ and\ \citenamefont {Gotor}}]{chicardi2020synthesis}%
  \BibitemOpen
  \bibfield  {author} {\bibinfo {author} {\bibfnamefont {E.}~\bibnamefont {Chicardi}}, \bibinfo {author} {\bibfnamefont {C.}~\bibnamefont {Garc{\'\i}a-Garrido}}, \bibinfo {author} {\bibfnamefont {J.}~\bibnamefont {Hern{\'a}ndez-Saz}},\ and\ \bibinfo {author} {\bibfnamefont {F.}~\bibnamefont {Gotor}},\ }\bibfield  {title} {\bibinfo {title} {Synthesis of all equiatomic five-transition metals high entropy carbides of the ivb (ti, zr, hf) and vb (v, nb, ta) groups by a low temperature route},\ }\href@noop {} {\bibfield  {journal} {\bibinfo  {journal} {Ceramics International}\ }\textbf {\bibinfo {volume} {46}},\ \bibinfo {pages} {21421} (\bibinfo {year} {2020})}\BibitemShut {NoStop}%
\bibitem [{\citenamefont {Raju~Natarajan}\ and\ \citenamefont {Van~der Ven}(2019)}]{raju_natarajan_toward_2019}%
  \BibitemOpen
  \bibfield  {author} {\bibinfo {author} {\bibfnamefont {A.}~\bibnamefont {Raju~Natarajan}}\ and\ \bibinfo {author} {\bibfnamefont {A.}~\bibnamefont {Van~der Ven}},\ }\bibfield  {title} {\bibinfo {title} {Toward an {Understanding} of {Deformation} {Mechanisms} in {Metallic} {Lithium} and {Sodium} from {First}-{Principles}},\ }\href@noop {} {\bibfield  {journal} {\bibinfo  {journal} {Chemistry of Materials}\ }\textbf {\bibinfo {volume} {31}},\ \bibinfo {pages} {8222} (\bibinfo {year} {2019})}\BibitemShut {NoStop}%
\bibitem [{\citenamefont {Baruffi}\ \emph {et~al.}(2023)\citenamefont {Baruffi}, \citenamefont {Ghazisaeidi}, \citenamefont {Rodney},\ and\ \citenamefont {Curtin}}]{BARUFFI2023115536}%
  \BibitemOpen
  \bibfield  {author} {\bibinfo {author} {\bibfnamefont {C.}~\bibnamefont {Baruffi}}, \bibinfo {author} {\bibfnamefont {M.}~\bibnamefont {Ghazisaeidi}}, \bibinfo {author} {\bibfnamefont {D.}~\bibnamefont {Rodney}},\ and\ \bibinfo {author} {\bibfnamefont {W.}~\bibnamefont {Curtin}},\ }\bibfield  {title} {\bibinfo {title} {Equilibrium versus non-equilibrium stacking fault widths in nicocr},\ }\href@noop {} {\bibfield  {journal} {\bibinfo  {journal} {Scripta Materialia}\ }\textbf {\bibinfo {volume} {235}},\ \bibinfo {pages} {115536} (\bibinfo {year} {2023})}\BibitemShut {NoStop}%
\bibitem [{\citenamefont {Li}\ \emph {et~al.}(2024)\citenamefont {Li}, \citenamefont {Zhao}, \citenamefont {Jiang}, \citenamefont {Cao}, \citenamefont {Xiao}, \citenamefont {Song}, \citenamefont {Hong}, \citenamefont {Gou},\ and\ \citenamefont {Liang}}]{LI2024174091}%
  \BibitemOpen
  \bibfield  {author} {\bibinfo {author} {\bibfnamefont {P.}~\bibnamefont {Li}}, \bibinfo {author} {\bibfnamefont {C.}~\bibnamefont {Zhao}}, \bibinfo {author} {\bibfnamefont {Y.}~\bibnamefont {Jiang}}, \bibinfo {author} {\bibfnamefont {F.}~\bibnamefont {Cao}}, \bibinfo {author} {\bibfnamefont {P.}~\bibnamefont {Xiao}}, \bibinfo {author} {\bibfnamefont {Y.}~\bibnamefont {Song}}, \bibinfo {author} {\bibfnamefont {Z.}~\bibnamefont {Hong}}, \bibinfo {author} {\bibfnamefont {S.}~\bibnamefont {Gou}},\ and\ \bibinfo {author} {\bibfnamefont {S.}~\bibnamefont {Liang}},\ }\bibfield  {title} {\bibinfo {title} {The relationship between deformation mechanisms and mechanical properties in nanocrystalline cu/ag-bilayer alloy},\ }\href@noop {} {\bibfield  {journal} {\bibinfo  {journal} {Journal of Alloys and Compounds}\ }\textbf {\bibinfo {volume} {986}},\ \bibinfo {pages} {174091} (\bibinfo {year} {2024})}\BibitemShut {NoStop}%
\bibitem [{\citenamefont {Kaufmann}\ \emph {et~al.}(2020)\citenamefont {Kaufmann}, \citenamefont {Maryanovsky}, \citenamefont {Mellor}, \citenamefont {Zhu}, \citenamefont {Rosengarten}, \citenamefont {Harrington}, \citenamefont {Oses}, \citenamefont {Toher}, \citenamefont {Curtarolo},\ and\ \citenamefont {Vecchio}}]{kaufmann_discovery_2020}%
  \BibitemOpen
  \bibfield  {author} {\bibinfo {author} {\bibfnamefont {K.}~\bibnamefont {Kaufmann}}, \bibinfo {author} {\bibfnamefont {D.}~\bibnamefont {Maryanovsky}}, \bibinfo {author} {\bibfnamefont {W.~M.}\ \bibnamefont {Mellor}}, \bibinfo {author} {\bibfnamefont {C.}~\bibnamefont {Zhu}}, \bibinfo {author} {\bibfnamefont {A.~S.}\ \bibnamefont {Rosengarten}}, \bibinfo {author} {\bibfnamefont {T.~J.}\ \bibnamefont {Harrington}}, \bibinfo {author} {\bibfnamefont {C.}~\bibnamefont {Oses}}, \bibinfo {author} {\bibfnamefont {C.}~\bibnamefont {Toher}}, \bibinfo {author} {\bibfnamefont {S.}~\bibnamefont {Curtarolo}},\ and\ \bibinfo {author} {\bibfnamefont {K.~S.}\ \bibnamefont {Vecchio}},\ }\bibfield  {title} {\bibinfo {title} {Discovery of high-entropy ceramics via machine learning},\ }\href@noop {} {\bibfield  {journal} {\bibinfo  {journal} {npj Computational Materials}\ }\textbf {\bibinfo {volume} {6}},\ \bibinfo {pages} {42} (\bibinfo {year} {2020})}\BibitemShut {NoStop}%
\bibitem [{\citenamefont {Varvenne}\ \emph {et~al.}(2016)\citenamefont {Varvenne}, \citenamefont {Luque},\ and\ \citenamefont {Curtin}}]{VARVENNE2016164}%
  \BibitemOpen
  \bibfield  {author} {\bibinfo {author} {\bibfnamefont {C.}~\bibnamefont {Varvenne}}, \bibinfo {author} {\bibfnamefont {A.}~\bibnamefont {Luque}},\ and\ \bibinfo {author} {\bibfnamefont {W.~A.}\ \bibnamefont {Curtin}},\ }\bibfield  {title} {\bibinfo {title} {Theory of strengthening in fcc high entropy alloys},\ }\href@noop {} {\bibfield  {journal} {\bibinfo  {journal} {Acta Materialia}\ }\textbf {\bibinfo {volume} {118}},\ \bibinfo {pages} {164} (\bibinfo {year} {2016})}\BibitemShut {NoStop}%
\bibitem [{\citenamefont {Varvenne}\ \emph {et~al.}(2017)\citenamefont {Varvenne}, \citenamefont {Leyson}, \citenamefont {Ghazisaeidi},\ and\ \citenamefont {Curtin}}]{VARVENNE2017660}%
  \BibitemOpen
  \bibfield  {author} {\bibinfo {author} {\bibfnamefont {C.}~\bibnamefont {Varvenne}}, \bibinfo {author} {\bibfnamefont {G.}~\bibnamefont {Leyson}}, \bibinfo {author} {\bibfnamefont {M.}~\bibnamefont {Ghazisaeidi}},\ and\ \bibinfo {author} {\bibfnamefont {W.}~\bibnamefont {Curtin}},\ }\bibfield  {title} {\bibinfo {title} {Solute strengthening in random alloys},\ }\href@noop {} {\bibfield  {journal} {\bibinfo  {journal} {Acta Materialia}\ }\textbf {\bibinfo {volume} {124}},\ \bibinfo {pages} {660} (\bibinfo {year} {2017})}\BibitemShut {NoStop}%
\bibitem [{\citenamefont {Liu}\ and\ \citenamefont {Curtin}(2024)}]{LIU2024119471}%
  \BibitemOpen
  \bibfield  {author} {\bibinfo {author} {\bibfnamefont {X.}~\bibnamefont {Liu}}\ and\ \bibinfo {author} {\bibfnamefont {W.}~\bibnamefont {Curtin}},\ }\bibfield  {title} {\bibinfo {title} {Atomistic simulations reveal strength reductions due to short-range order in alloys},\ }\href@noop {} {\bibfield  {journal} {\bibinfo  {journal} {Acta Materialia}\ }\textbf {\bibinfo {volume} {263}},\ \bibinfo {pages} {119471} (\bibinfo {year} {2024})}\BibitemShut {NoStop}%
\bibitem [{\citenamefont {Maresca}\ and\ \citenamefont {Curtin}(2020{\natexlab{a}})}]{MARESCA2020144}%
  \BibitemOpen
  \bibfield  {author} {\bibinfo {author} {\bibfnamefont {F.}~\bibnamefont {Maresca}}\ and\ \bibinfo {author} {\bibfnamefont {W.~A.}\ \bibnamefont {Curtin}},\ }\bibfield  {title} {\bibinfo {title} {Theory of screw dislocation strengthening in random bcc alloys from dilute to ``high-entropy'' alloys},\ }\href@noop {} {\bibfield  {journal} {\bibinfo  {journal} {Acta Materialia}\ }\textbf {\bibinfo {volume} {182}},\ \bibinfo {pages} {144} (\bibinfo {year} {2020}{\natexlab{a}})}\BibitemShut {NoStop}%
\bibitem [{\citenamefont {Maresca}\ and\ \citenamefont {Curtin}(2020{\natexlab{b}})}]{MARESCA2020235}%
  \BibitemOpen
  \bibfield  {author} {\bibinfo {author} {\bibfnamefont {F.}~\bibnamefont {Maresca}}\ and\ \bibinfo {author} {\bibfnamefont {W.~A.}\ \bibnamefont {Curtin}},\ }\bibfield  {title} {\bibinfo {title} {Mechanistic origin of high strength in refractory bcc high entropy alloys up to 1900k},\ }\href@noop {} {\bibfield  {journal} {\bibinfo  {journal} {Acta Materialia}\ }\textbf {\bibinfo {volume} {182}},\ \bibinfo {pages} {235} (\bibinfo {year} {2020}{\natexlab{b}})}\BibitemShut {NoStop}%
\bibitem [{\citenamefont {Karumuri}\ \emph {et~al.}(2025)\citenamefont {Karumuri}, \citenamefont {Hernandez}, \citenamefont {Mishra}, \citenamefont {McClure}, \citenamefont {Tucker}, \citenamefont {Flanagan}, \citenamefont {Hwang}, \citenamefont {Sandhage}, \citenamefont {Bilionis}, \citenamefont {Titus},\ and\ \citenamefont {Strachan}}]{Karumuri2025}%
  \BibitemOpen
  \bibfield  {author} {\bibinfo {author} {\bibfnamefont {S.}~\bibnamefont {Karumuri}}, \bibinfo {author} {\bibfnamefont {A.}~\bibnamefont {Hernandez}}, \bibinfo {author} {\bibfnamefont {S.}~\bibnamefont {Mishra}}, \bibinfo {author} {\bibfnamefont {Z.}~\bibnamefont {McClure}}, \bibinfo {author} {\bibfnamefont {V.}~\bibnamefont {Tucker}}, \bibinfo {author} {\bibfnamefont {J.~C.}\ \bibnamefont {Flanagan}}, \bibinfo {author} {\bibfnamefont {S.}~\bibnamefont {Hwang}}, \bibinfo {author} {\bibfnamefont {K.~H.}\ \bibnamefont {Sandhage}}, \bibinfo {author} {\bibfnamefont {I.}~\bibnamefont {Bilionis}}, \bibinfo {author} {\bibfnamefont {M.~S.}\ \bibnamefont {Titus}},\ and\ \bibinfo {author} {\bibfnamefont {A.}~\bibnamefont {Strachan}},\ }\bibfield  {title} {\bibinfo {title} {Design of high-hardness complex concentrated alloys from physics, machine learning, and experiments},\ }\href@noop {} {\bibfield  {journal} {\bibinfo  {journal} {Journal of Applied Physics}\ }\textbf {\bibinfo {volume} {138}},\ \bibinfo {pages}
  {085106} (\bibinfo {year} {2025})}\BibitemShut {NoStop}%
\end{thebibliography}%
\end{document}


\title{Supplementary Materials for ``Synthesizability, hardness, and stacking order in multicomponent transition metal carbides from machine-learned potentials''}

\author{Xin Liu}
\email{xin.liu@epfl.ch}
\affiliation{Laboratory of materials design and simulation (MADES), Institute of Materials, \'{E}cole Polytechnique F\'{e}d\'{e}rale de Lausanne}

\author{Anirudh Raju Natarajan}
\email{anirudh.natarajan@epfl.ch}
\affiliation{Laboratory of materials design and simulation (MADES), Institute of Materials, \'{E}cole Polytechnique F\'{e}d\'{e}rale de Lausanne}
\affiliation{National Centre for Computational Design and Discovery of Novel Materials (MARVEL), \'{E}cole Polytechnique F\'{e}d\'{e}rale de Lausanne}

\maketitle

\section{MACE Model Performance}
\label{subsec:appendix:hull}

\begin{figure}[H]
  \centering
  \includegraphics[width=0.8\textwidth]{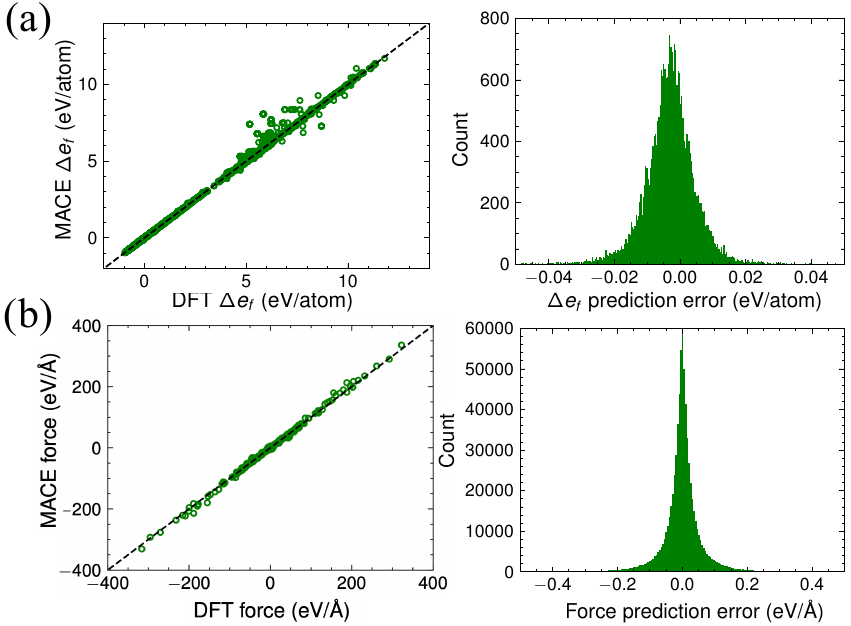}
  \caption{Prediction accuracy of the \texttt{MACE-100} model compared with DFT. (a) Parity plot and error distribution for formation energies. (b) Parity plot and error distribution for atomic forces.}
  \label{fig:energy_force}
\end{figure}

\Cref{fig:energy_force} shows the prediction performance of the fine-tuned model (\texttt{MACE-100}) for formation energies and atomic forces. When evaluated over the full dataset, the root-mean-square errors (RMSEs) are 92 meV/atom for energy and 135 meV/\AA\ for force. Restricting to the low-energy subset (formation energy $<$ 0.5 eV/atom), the errors drop significantly to 8 meV/atom and 51 meV/\AA, respectively.

\begin{figure}[H]
  \centering
  \includegraphics[width=0.8\textwidth]{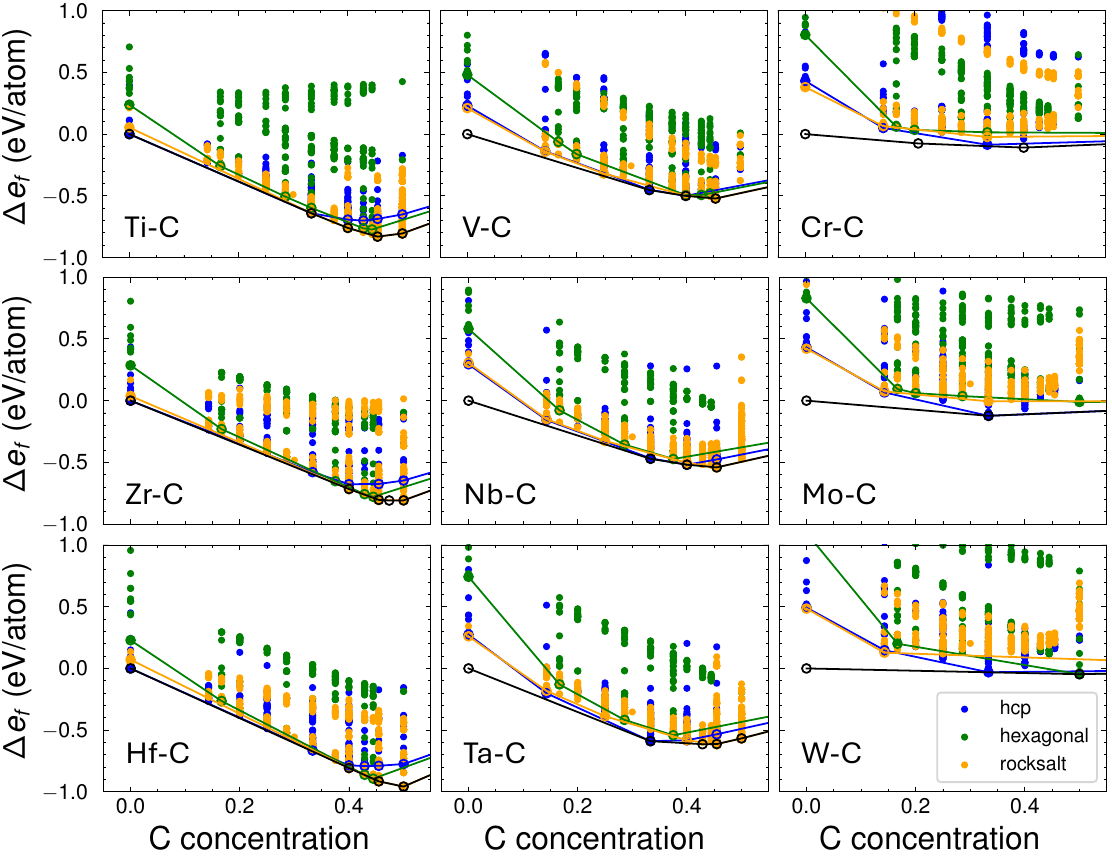}
 \caption{Binary carbide formation energy hulls computed with the \texttt{MACE-MPA-0} foundation model. Points are colored by structural prototype: hcp (blue), hexagonal (green), and rocksalt (orange). The black line denotes the global convex hull and open circles indicate on-hull structures.}
  \label{fig:formation_hull_mpa_0}
\end{figure}

\begin{figure}[H]
  \centering
  \includegraphics[width=0.8\textwidth]{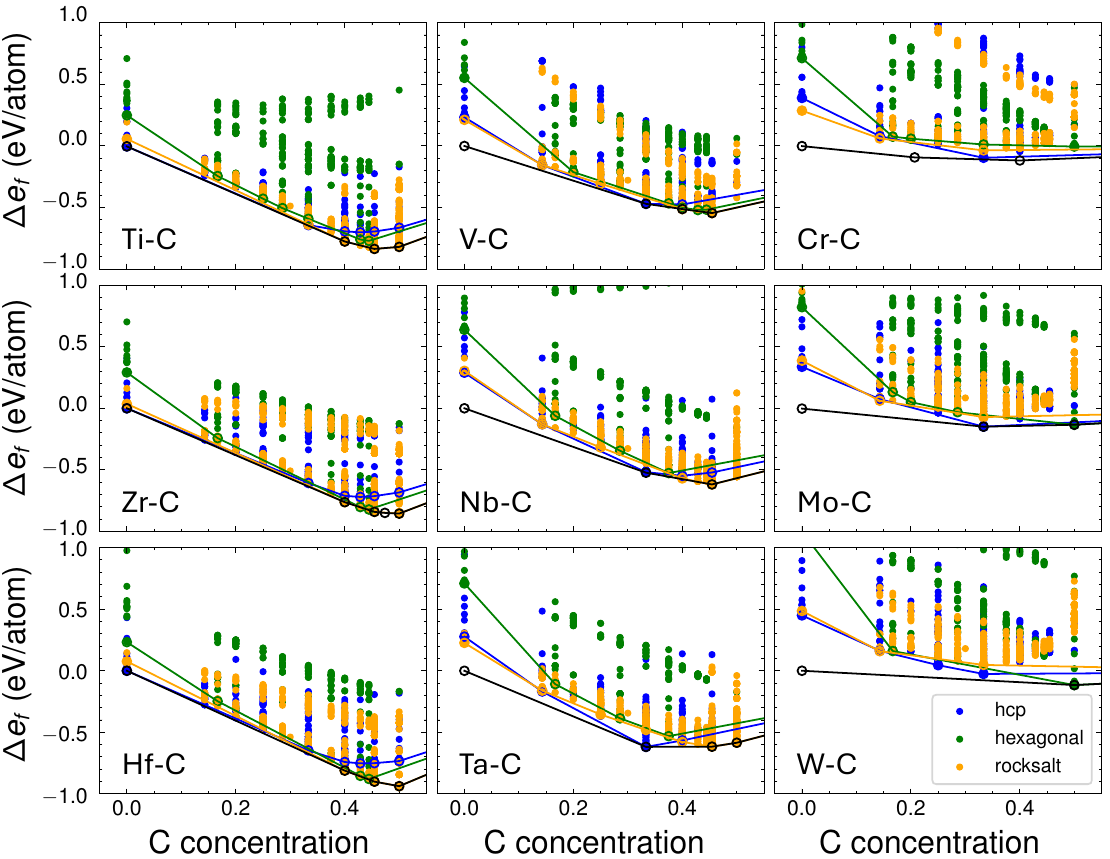}
  \caption{Binary carbide formation energy hulls computed with the partially fine-tuned \texttt{MACE-20} model. Color and symbol conventions follow \cref{fig:formation_hull_mpa_0}.}
  \label{fig:formation_hull_mace20}
\end{figure}

\Cref{fig:formation_hull_mpa_0,fig:formation_hull_mace20} show the binary carbide formation energy hulls predicted by the foundation model (\texttt{MACE-MPA-0}) and the partially fine-tuned model (\texttt{MACE-20}). As observed in the main text (Figure 5), \texttt{MACE-20} predictions closely track those of the fully trained \texttt{MACE-100} and are largely consistent with DFT. The foundation model shows reasonable accuracy but tends to deviate near phase boundaries.

\section{Hardness Prediction Models}
\label{subsec:appendix:Hv}

The following models are used to estimate Vickers hardness $H_v$ from elastic properties:

\begin{equation}
    H_{v}^{\mathrm{Teter}} = 0.151\, G
\label{eq:Hv_teter}
\end{equation}

\begin{equation}
    H_{v}^{\mathrm{Chen}} = 2 \left(\frac{G^3}{B^2}\right)^{0.585} - 3
\label{eq:Hv_chen}
\end{equation}

\begin{equation}
    H_{v}^{\mathrm{Tian}} = 0.92 \left(\frac{G}{B}\right)^{1.137} G^{0.708}
\label{eq:Hv_tian}
\end{equation}

\begin{equation}
    H_{v}^{\mathrm{Mazhnik}} = 0.096 \,\frac{1 - 8.5\nu + 19.5\nu^2}{1 - 7.5\nu + 12.2\nu^2 + 19.6\nu^3}\, E
\label{eq:Hv_mazhnik}
\end{equation}

Here, $G$ is the shear modulus, $B$ is the bulk modulus, $E$ is Young’s modulus, and $\nu$ is Poisson’s ratio.

\begin{figure}[H]
  \centering
  \includegraphics[width=0.8\textwidth]{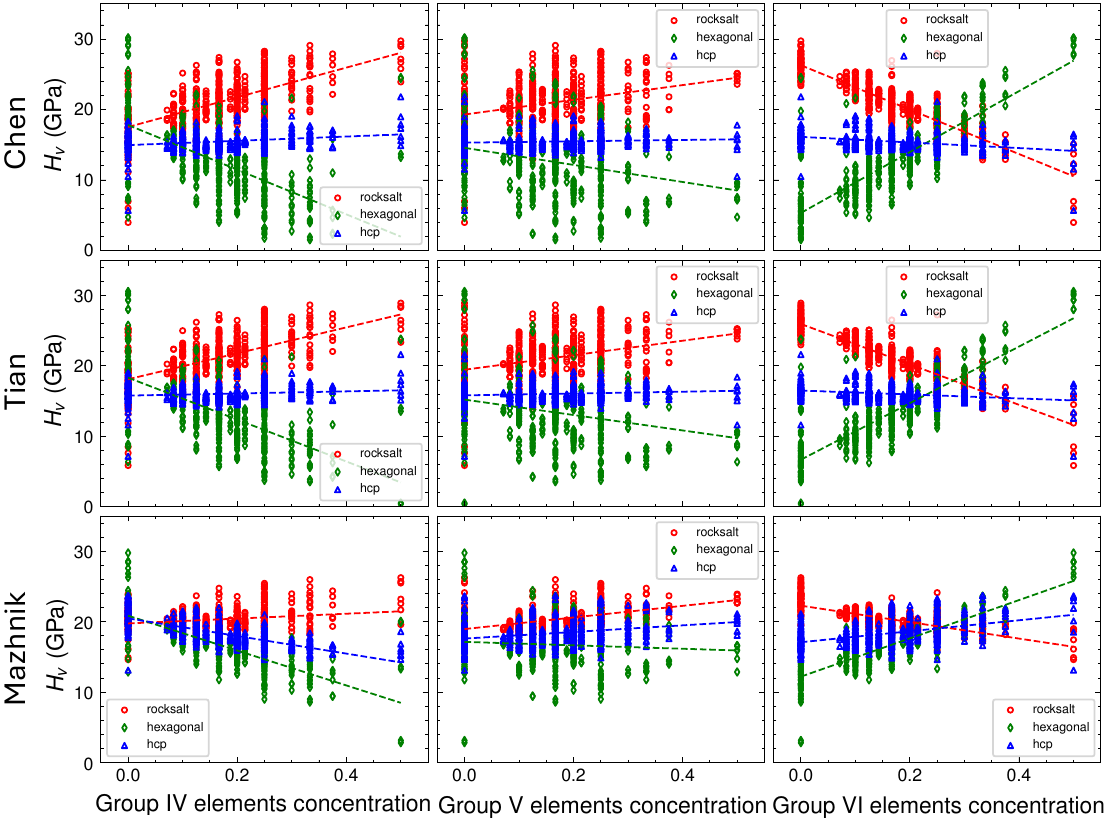}
  \caption{Vickers hardness $H_v$ for equiatomic transition metal carbides as a function of the fraction of metals from groups 4, 5, and 6, computed with the Chen, Tian, and Mazhnik surrogate models. Results are shown separately for the rocksalt, hcp, and hexagonal prototypes. Dashed lines are linear fits.}
  \label{fig:Hv_models}
\end{figure}

\Cref{fig:Hv_models} shows how predicted hardness varies with the fraction of Group 4,5 and 6 elements in equiatomic SQS carbides. Across all three structure types (rocksalt, hcp, hexagonal), the Chen, Tian, and Mazhnik models reproduce the same qualitative trends observed with Teter’s model. For rocksalt phases, $H_v$ increases with the fraction of Group 4 and 5 elements, while hexagonal structures show higher hardness with increased Group 6 content. The hcp prototype shows only mild variation across group fractions.

\begin{figure}[H]
  \centering
  \includegraphics[width=0.8\textwidth]{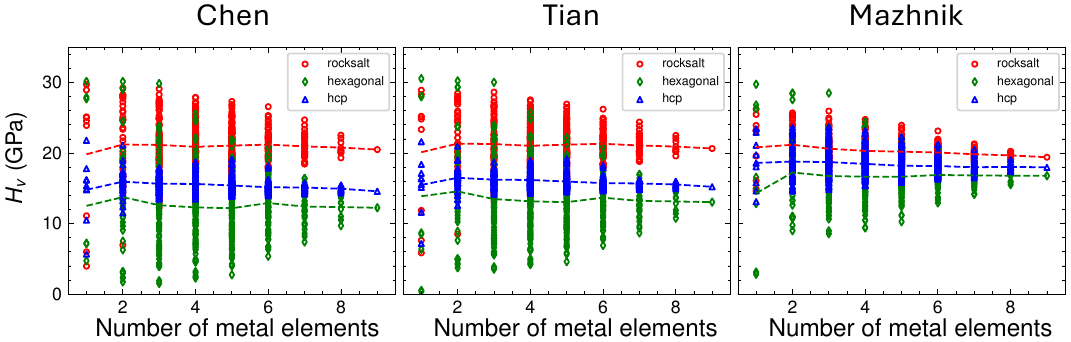}
  \caption{Vickers hardness $H_v$ as a function of the number of constituent metals for equiatomic transition metal carbides, computed with the Chen, Tian, and Mazhnik surrogate models. Dashed lines indicate the average hardness for each metal count.}
  \label{fig:Hv_metal_n_models}
\end{figure}